\documentclass[iop,apj]{emulateapj}
\usepackage{amsbsy}
\usepackage{amsfonts}
\usepackage{amssymb}
\usepackage{bm,ulem}
\usepackage{mathrsfs}
\usepackage{pifont}
\usepackage{stmaryrd}
\usepackage{textcomp,bookmark}
\usepackage{portland,xspace}
\usepackage{amsmath,amsxtra}
\usepackage[OT2,OT1]{fontenc}
\usepackage{graphicx}
\usepackage{hyperref}
\usepackage{natbib}
\usepackage{array,colortbl,diagbox}
\usepackage[table]{xcolor}
\newcommand{\msun}{\ifmmode {{M}_\odot} \else  {${M}_\odot$} \fi}
\newcommand{\hms}{\ifmmode {h^{-1}{M}_\odot} \else  {$h^{-1}{M}_\odot$} \fi}
\newcommand{\mpc}{\ifmmode {{\rm Mpc}} \else  {${\rm Mpc}$} \fi}
\newcommand{\kpc}{\ifmmode {{\rm kpc}} \else  {${\rm kpc}$} \fi}
\newcommand{\hmpc}{\ifmmode {h^{-1}{\rm Mpc}} \else  {$h^{-1}{\rm Mpc}$} \fi}
\newcommand{\hkpc}{\ifmmode {h^{-1}{\rm kpc}} \else  {$h^{-1}{\rm kpc}$} \fi}
\newcommand{\amin}{\ifmmode {{\rm arcmin}} \else  {${\rm arcmin}$} \fi}
\newcommand{\mv}{\ifmmode {{M}_{\rm vir}} \else  {${M}_{\rm vir}$} \fi}
\newcommand{\mt}{\ifmmode {{M}_{200}} \else  {${M}_{200}$} \fi}
\newcommand{\mlow}{\ifmmode {{M}_{\rm low}} \else  {${M}_{\rm low}$} \fi}
\newcommand{\neff}{\ifmmode {{n}_{\rm eff}} \else  {${n}_{\rm eff}$} \fi}
\newcommand{\cv}{\ifmmode {{c}_{\rm vir}} \else  {${c}_{\rm vir}$} \fi}
\newcommand{\ct}{\ifmmode {{c}_{200}} \else  {${c}_{200}$} \fi}
\newcommand{\rv}{\ifmmode {{r}_{\rm vir}} \else  {${r}_{\rm vir}$} \fi}
\newcommand{\rt}{\ifmmode {{r}_{200}} \else  {${r}_{200}$} \fi}
\newcommand{\rdlt}{\ifmmode {{r}_{\scriptscriptstyle\Delta}} \else  {${r}_{\scriptscriptstyle\Delta}$} \fi}
\newcommand{\mdlt}{\ifmmode {{M}_{\scriptscriptstyle\Delta}} \else  {${M}_{\scriptscriptstyle\Delta}$} \fi}
\newcommand{\cdlt}{\ifmmode {{c}_{\scriptscriptstyle\Delta}} \else  {${c}_{\scriptscriptstyle\Delta}$} \fi}
\newcommand{\smM}{\ifmmode {\scriptscriptstyle M} \else  {${\scriptscriptstyle M}$} \fi}
\newcommand{\smT}{\ifmmode {\scriptscriptstyle T} \else  {${\scriptscriptstyle T}$} \fi}
\newcommand{\zbpz}{\ifmmode {z_{\scriptscriptstyle {\rm BPZ}}} \else  {${z_{\scriptscriptstyle {\rm BPZ}}}$} \fi}
\newcommand{\hh}{\ifmmode {h^{-1}} \else  {$h^{-1}$} \fi}
\newcommand{\vth}{\ifmmode {\vec{\theta}} \else  {$\vec{\theta}$} \fi}
\newcommand{\ie}{i.e.,}
\newcommand{\eg}{e.g.,}

\newcommand{\lgg}{\ifmmode {{\rm log}} \else  {${\rm log}$} \fi}
\newcommand{\sn}{\ifmmode {\sigma_n} \else  {$\sigma_n$} \fi}
\newcommand{\snsq}{\ifmmode {\sigma_n^2} \else  {$\sigma_n^2$} \fi}

\begin{document}

\title{Mass--concentration relation of clusters of galaxies from CFHTLenS}
\author{Wei Du$^{1,2}$, Zuhui Fan$^{2,3}$, Huanyuan Shan$^4$, Gong-Bo Zhao$^1$, Giovanni Covone$^{5,6}$, Liping Fu$^7$, Jean-Paul Kneib$^{4,8}$}
\affil{$^1$National Astronomical Observatories, Chinese Academy of Science, Beijing 100012, China; duwei@bao.ac.cn\\
$^2$Department of Astronomy, Peking University, Beijing 100871, China; fanzuhui@pku.edu.cn\\
$^3$Collaborative Innovation Center of Modern Astronomy and Space Exploration, Nanjing, 210093, China\\
$^4$Laboratoire d'astrophysique (LASTRO), Ecole Polytechnique F\'ed\'erale de Lausanne (EPFL), Observatoire de Sauverny,
CH-1290 Versoix, Switzerland\\
$^5$ Dipartimento di Fisica, Universit`a di Napoli Federico II, Via Cinthia, I-80126 Napoli, Italy\\
$^6$ INFN Sez. di Napoli, Compl. Univ. Monte S. Angelo, Via Cinthia, I-80126 Napoli, Italy \\
$^7$ Shanghai Key Lab for Astrophysics, Shanghai Normal University, Shanghai 200234, China\\
$^8$ Aix Marseille Universit\'e, CNRS, LAM (Laboratoire d'Astrophysique de Marseille) UMR 7326, 13388, Marseille, France}

\begin{abstract}
Based on weak lensing data from the Canada-France-Hawaii Telescope Lensing Survey (CFHTLenS), in this paper we study the mass--concentration ($M$--$c$) relation for $\sim 200$
redMaPPer clusters in the fields.
We extract the $M$--$c$ relation by measuring the density profiles of individual clusters instead of using stacked weak lensing signals.
By performing Monte Carlo simulations, we demonstrate that although the signal-to-noise ratio for each individual cluster is low,
the unbiased $M$--$c$ relation can still be reliably derived from a large sample of clusters
by carefully taking into account the impacts of shape noise, cluster center offset, dilution effect from member or foreground galaxies,
and the projection effect. Our results show that within error bars the derived $M$--$c$ relation for redMaPPer clusters
is in agreement with simulation predictions. There is a weak deviation in that the halo concentrations calibrated by Monte Carlo simulations
are somewhat higher than that predicted from ${\it Planck}$ cosmology.
\end{abstract}


\section{Introduction}
Clusters of galaxies are important tracers of large-scale structures in the universe.
Their properties, such as the mass function and density profiles, are tightly related to cosmology and the nature of dark matter particles
and therefore carry abundant cosmological information
\citep[\eg][]{1974ApJ...187..425P,2004MNRAS.353..457A,2004ApJ...606..819M,2008ApJ...679.1173R,2008ApJ...688..709T,2011ApJS..192...18K,
2012ApJ...758..128C,2012ApJ...748..120D,2013A&A...555A..66L,2014A&A...571A..16P}.

In the cold dark matter (CDM) paradigm, high-resolution simulations reveal an approximately universal density profile for dark matter halos
with the logarithmic slope ${\ln\rho(r)}/{\ln r}$ $\sim -1$ at the inner part and $\sim -3$
at the outer part \citep{1990ApJ...356..359H,1996ApJ...462..563N,1997ApJ...490..493N,1999ApJ...517...64H,2014ApJ...789....1D}.
A characteristic radius $r_s$ for a halo is conventionally defined as the radius at which ${\ln\rho(r)}/{\ln r}=-2$.
The corresponding concentration parameter is $\cdlt=\rdlt/r_s$, where $\rdlt$ is the halo radius within which the average density
is $\Delta$ times the critical density $\rho_{\rm crit}$ or the mean density of the universe, depending on specific definitions.
The larger $\cdlt$ is, the more centrally concentrated is the matter distribution.

The Navarro--Frenk--White (NFW; \citealt{1996ApJ...462..563N,1997ApJ...490..493N}) profile is the one widely adopted
to describe the density profile of dark matter halos, which contains two parameters, the characteristic density and radius.
The Einasto profile involving three parameters has also been proposed, aiming to give a better fit to the halo density profile
\citep{1965TrAlm...5...87E,2008MNRAS.387..536G,2012A&A...540A..70R,2014MNRAS.441.3359D}. In this paper, we will focus on
the NFW profile, which is given by
\begin{equation}\label{eq:rho}
\rho(r)=\frac{\rho_s}{(r/r_s)(1+r/r_s)^2},
\end{equation}
where $\rho_s$ and $r_s$ are respectively the characteristic density and scale radius. Equivalently, the two parameters can be replaced
by the halo mass $\mdlt$ within $\rdlt$ and the concentration parameter $\cdlt$. Simulations show that
there is a tight correlation between $\cdlt$ and $\mdlt$ with a scatter of $\sim 0.1-\sim 0.2$ in $\log \cdlt$ given an $\mdlt$
\citep{2000ApJ...535...30J,2001MNRAS.321..559B,2007MNRAS.381.1450N}. For the overdensity parameter $\Delta$,
different values can be taken for different purposes. For X-ray-related studies, $\Delta=500$ or even larger is usually adopted
because the X-ray emission of a cluster is strongly centrally concentrated \citep[\eg][]{2005A&A...441..893A,2007MNRAS.379..518M,2011MNRAS.413..691Y}.
For weak lensing analyses that can cover outer parts of clusters, $\Delta=200$ or $\Delta_{\rm vir}(z)$ is often used.
In this paper, we adopt $\Delta=200$ relative to the critical density of the universe, and we will drop the subscript $\Delta$ hereafter.

The correlation between the concentration parameter and the mass of dark matter halos, often referred to as the
mass--concentration ($M$--$c$) relation, carries important cosmological information.
Understanding its behaviors is therefore a subject under intensive investigations both theoretically via simulations and observationally.
From the theoretical side, while it is known that the $M$--$c$ relation is
related closely to the halo assembly history, the predictions from different simulation studies still vary
significantly \citep[e.g.,][]{2011ApJ...740..102K,2014arXiv1411.4001K,2012MNRAS.427.1322L,2013MNRAS.432.1103L,2012MNRAS.423.3018P,
2014MNRAS.441.3359D,2015MNRAS.452.1217C}.
Recent investigations show that a more universal form related to the concentration parameter $c$ may be
the concentration--peak height $\nu=\delta_c(z)/\sigma(M,z)$ relation,
where $\delta_c(z)$ is the critical density fluctuation extrapolated linearly to the epoch of halo formation
and $\sigma(M,z)$ is the rms of linear density fluctuations at redshift $z$ smoothed over a kernel with the scale corresponding
to mass $M$ \citep[\eg][]{2014MNRAS.441..378L,2015ApJ...799..108D}.
This is expected physically given the mass accretion histories of dark matter halos.
The $M$--$c$ relation is then implicitly embedded in the $c$--$\nu$ relation. A more accurate dependence of $c$ on $M$, $\nu$, or other physical quantities
needs to be further studied for the purpose of precise cosmology. On the other hand, a power-law $M$--$c$ relation is a relatively good approximation for clusters spanning a narrow mass range.

Assuming $c\approx A (M/M_{\rm p})^\alpha$, simulations show that the slope $\alpha \sim -0.1$ at low redshift, which is nearly
independent of cosmology. However, the amplitude $A$ depends on cosmology and increases significantly with time
\citep[e.g.,][]{2001MNRAS.321..559B,2003ApJ...597L...9Z,2004A&A...416..853D,2006ApJ...646..815S,2008MNRAS.390L..64D,
2008MNRAS.387..536G,2009ApJ...707..354Z,2011MNRAS.411..584M,2012MNRAS.424.1244F,2013ApJ...766...32B,2014ApJ...785...57D}.
It should be noted that different choices of the pivot mass $M_{rm p}$ and of $\Delta$ used to calculate halo masses also lead to different $A$.

Observationally, extensive analyses have been done to probe the $M$--$c$ relation of dark matter halos.
A number of studies based on individual analyses for a sample of clusters
show a steeper dependence of $c$ on $M$ than that predicted by simulations \citep[e.g.,][]{2010A&A...524A..68E,2010PASJ...62..811O,2012MNRAS.420.3213O,2012ApJ...761....1W,2013MNRAS.434..878S,2015ApJ...806....4M,2015arXiv150704385U}.
\citet{2012MNRAS.424.1244F} investigated the baryonic cooling effect on the profile of dark matter halos.
They conclude that the steep slope of $\alpha$ can be reproduced by adjusting the baryon fraction according to halo mass.
However, feedback from active galactic nuclei and supernovae can compensate for the contraction effect of baryonic cooling
\citep{2010MNRAS.405.2161D,2010MNRAS.406..434M,2011MNRAS.416.2539K}. Therefore, the net effect of baryonic physics is still under debate.

On the other hand, it is pointed out that the sample variance, selection bias, and the narrow mass range considered may lead to
a bias for the observationally derived $M$--$c$ relation \citep{2008JCAP...08..006M,2010A&A...524A..68E,2015MNRAS.449.2024S}.
In particular, the error correlations between $M$ and $c$ can affect the analyses significantly.
\citet{2013MNRAS.436..503A} show that the constraints on $M_{200}$ and $c$ of a dark matter halo derived by combining the observed
Einstein radius and the mass estimate for the inner part $M_{500}$ from optical richness have highly correlated
scatters, which in turn can result in a steep $M$--$c$ relation resembling the degenerate direction between $M$ and $c$.
Concerning weak lensing analyses, in \citet{2014ApJ...785...57D} with simulated clusters, we demonstrate that the shape noise
from background galaxies can lead to significant scatters for the constraints on $M$ and $c$.
The scatters are strongly correlated because of the degeneracy between $M$ and $c$ with respect to the weak lensing
shear signals. Such correlated scatters can bias the apparent $M$--$c$ relation and make it steeper than that of simulated halos.
Assuming the errors of $\log M$ and $\log c$ being Gaussian, we develop a Bayesian method taking into account the error correlations.
We demonstrate that for the number density of background galaxies $n_g>10\hbox{ arcmin}^{-2}$, our Bayesian method works well
in extracting the unbiased $M$--$c$ relation for dark matter halos \citep{2014ApJ...785...57D}.

The error correlations have been taken into account in recent observational lensing studies.
With high-quality weak lensing data from Suprime-Cam mounted on the 8.2 m Subaru telescope,
\citet{2015arXiv150704493O} analyze $50$ X-ray luminous clusters.
Applying a Bayesian linear regression method \citep[\eg][]{2007ApJ...665.1489K} with the error correlations between $M$ and $c$ included,
they obtain an $M$--$c$ relation for the cluster sample with a slope $\alpha = -0.13^{+0.16}_{-0.17}$, consistent with simulations.
\citet{2015arXiv150704385U} perform studies for $16$ bright X-ray clusters using both strong and weak lensing (shear and magnification)
observations. With the same Bayesian linear regression method adding a log-normal distribution for the intrinsic cluster mass, they derive a constraint on the
slope parameter $\alpha=-0.44\pm 0.19$. In these analyses, the correlated errors in $\log M$ and $\log c$ are taken to be Gaussian,
which can be a good approximation given their high-quality data with $n_g>10\hbox{ arcmin}^{-2}$ for weak lensing observations.

In this paper, we study the $M$--$c$ relation from weak lensing analyses using the Canada-France-Hawaii Telescope Lensing Survey (CFHTLenS).
We perform individual studies for $\sim 200$ redMaPPer clusters \citep{2014ApJ...785..104R} and derive the $M$--$c$ relation from the sample.
We will see in Section 2.3 that for about one-half of our clusters, the number density of background galaxies used in weak lensing analyses is
low with $n_g < 8\hbox{ arcmin}^{-2}$. Therefore the errors for the derived $\log M$ and $\log c$ are large
and cannot be well described by Gaussian ellipses. Consequently, the $M$--$c$ relation obtained from the Bayesian method
including only the Gaussian correlated errors still suffers from a bias \citep{2014ApJ...785...57D}.
On the other hand, in \citet{2014ApJ...785...57D}, we point out that including an additional mass-related observable
independent of weak lensing effects in the analyses can help to break the degeneracy between $M$ and $c$,
and therefore to reduce the apparent bias on the weak-lensing-derived $M$--$c$ relation.
Indeed, stacked weak lensing analyses in which the binning of clusters or groups is based on either their optical richness
or their luminosity/stellar mass show that the slope of the derived  $M$--$c$ relations is consistent with that predicted from numerical simulations,
although discrepancies exist for the amplitude
\citep[\eg][]{2007arXiv0709.1159J,2008JCAP...08..006M,2014ApJ...784L..25C,2015arXiv150200313S}.

In view of the large errors in weak-lensing-determined $M$ and $c$ from CFHTLenS, in this paper we therefore take the approach to add external constraints
on $M$ from the optical richness given by the redMaPPer catalog to break the degeneracy between $M$ and $c$.
We also pay attention to the impacts of other effects on our weak lensing studies, such as the off-center and the dilution effects.
We show that although the signal-to-noise ratio is low for individual clusters, an unbiased $M$--$c$ relation
can still be extracted from a large sample of clusters once the possible bias effects are taken into account carefully.
Throughout the paper, we adopt the cosmological model from {\it Planck} with $\Omega_M=0.315$, $\Omega_\Lambda=0.685$ and
$h=0.673$ \citep{2014A&A...571A..16P}.

The rest of the paper is organized as follows. In Section 2, we describe the data used in our studies, both the
foreground cluster catalog and the background galaxies. In Section 3, we present the procedures of weak lensing analyses.
The Monte Carlo simulation studies are shown in Section 4. The results are presented in Section 5.
Summaries will be given in Section 6.

\section{Observational Data}

\subsection{Foreground Clusters}

Our cluster sample is taken from the public redMaPPer v5.10 cluster catalog, which is based on the photometric data of Sloan Digital Sky Survey (SDSS) DR8
\citep{2011ApJS..193...29A,2014ApJ...785..104R,2015MNRAS.453...38R}. The redMaPPer algorithm is designed to find galaxy clusters using multicolor
photometric data through two stages of analyses. A calibration stage is meant to calibrate the model of red sequence galaxies
as a function of redshift empirically, which is then utilized in the cluster-finding stage to identify galaxy clusters and
measure their richness. The two stages proceed iteratively to give rise to a final cluster catalog.

For each redMaPPer cluster, an optimized richness $\lambda$ is obtained from the sum of membership probabilities, taking into account
the masked region and survey depth. The cluster catalog contains clusters with at least $20$ detected members, i.e., with $\lambda/S(z)> 20$,
where $S(z)$ is the richness scale factor accounting for the limited survey depth \citep{2014ApJ...785..104R,2015MNRAS.453...38R}.
The photometric redshift (photo-$z$) distribution of the clusters ranges from $z_\lambda= 0.08$ to 0.55.
For cluster centering, the redMaPPer algorithm assigns each member galaxy a centering probability $P_{\rm cen}$
according to its luminosity, its red sequence redshift, and the local cluster galaxy density around it.
The probabilities that the other member galaxies are central are also taken into consideration.
In our analyses, we take the galaxy with the highest $P_{\rm cen}$ as the central galaxy of a cluster.
It will be seen later that the information of $P_{\rm cen}$ is very helpful in our weak lensing analyses.

The overlapping area between SDSS DR8 and CFHTLenS is $\sim 120\ \deg^2$.
In the overlapping regions, there are $287$ redMaPPer clusters. They form the base sample used in our studies.

Because lensing effects depend on line-of-sight (LOS) mass distributions, we need to treat properly the cases with
more than one cluster along an LOS. For that, we divide the clusters into isolated ones and paired ones.
Two clusters are defined as a pair if the angular separation $\theta_{12}$ between their central galaxies is less than
$1.5\hmpc[1/D_d(z_1)+1/D_d(z_2)]~(z_1<z_2)$, where $D_d(z_1)$ and $D_d(z_2)$ are the angular diameter distances to the two clusters, respectively.
In accord with $\theta_{12}$, we define $R_{12}$ as the corresponding physical size at the lower-redshift lens plane with
$R_{12}=D_d(z_1)\theta_{12}$. In the current analyses, we discard cases with more than two clusters along an LOS.
Also, with the objective of reducing the mask effect on density profile fitting, we further exclude
clusters (and pairs) with mask fraction $f_{\rm m}>5\%$ within the radius of $0.5\hmpc/D_d$ around their centers.
Here we follow the method of \citet{2013ApJ...774..111S} to define masks. Specifically, a pixel with size $0.1~\amin$
is marked as a mask if there are no source galaxies from CFHTLenS within a radius of $0.5~\amin$ around it.
To avoid possible false masks due to sparsity of source galaxies, we include all CFHTLenS galaxies with photo $z$ when defining masks.
After the exclusion, our final lens sample contains a total of 220 clusters with 158 isolated ones and 31 pairs.

\begin{figure*}
  \centering
  \includegraphics[width=0.8\textwidth]{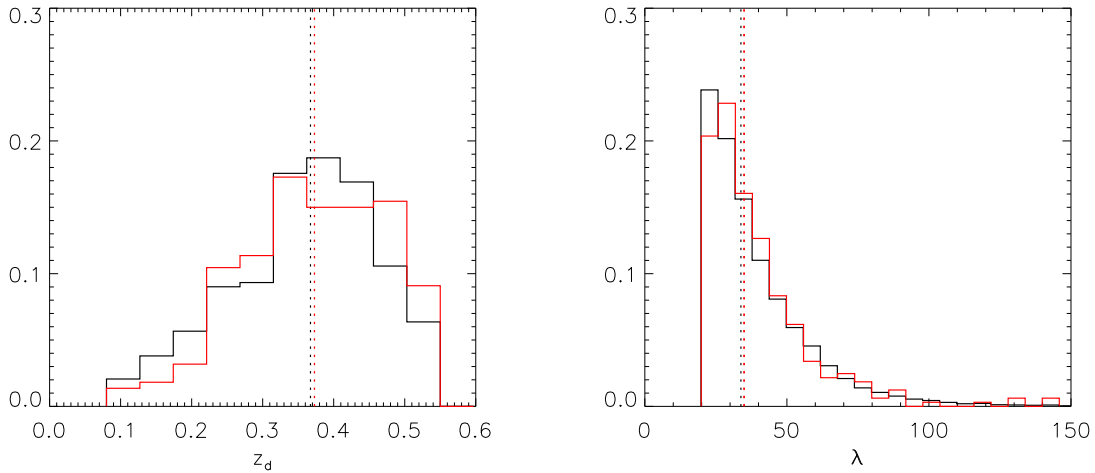}\\
  \caption{Normalized redshift (left) and richness (right) distribution of clusters. In both panels, black histograms are for all of the redMaPPer clusters (26,350 in total). Red is for the clusters in CFHTLenS. Dotted lines indicate the median values of the corresponding distributions.}\label{Fig:richz}
\end{figure*}

Figure \ref{Fig:richz} shows the redshift (left) and richness (right) distributions for our lens sample (red) and the
general redMaPPer catalog with $26,350$ clusters (black), respectively. The dotted lines indicate the corresponding median values.
It is seen that our lens sample is a fair representation of the full catalog.

\subsection{Source Galaxies}

Our weak lensing analyses are based on the CFHTLenS catalog, which is constructed from the five years of observational data from the CFHT Legacy Survey (CFHTLS)
\citep{2012MNRAS.427..146H,2012MNRAS.421.2355H,2013MNRAS.431.1547B,2013MNRAS.433.2545E,2013MNRAS.429.2858M}. The survey has four patches
(W1, W2, W3, and W4) and in total 171 pointings with complete five-band $u^{*}, g^\prime,r^\prime,i^\prime(y^\prime),z^\prime$ observations,
where the $y^\prime$-band filter replaced the original i-band filter after it was broken in 2008. In this paper, we only use the shape measurements
in the $y^\prime$ band for weak lensing studies if the source galaxies were observed by both $y^\prime$ and $i^\prime$ bands.
The ``lensfit'' method is employed for shape measurements of source galaxies. We select target galaxies by setting the object
classification parameter $\rm FITCLASS=0$. The mask parameter is set to $\rm MASK \le 1$ for selecting galaxies not
severely affected by various observational effects.  The lensfit weight for the selected galaxies satisfies $\omega>0$ \citep{2013MNRAS.429.2858M}.
For CFHTLenS, the photo-$z$ distribution $P(z)$ for each galaxy is estimated by the Bayesian Photometric Redshifts (BPZ) code based on the five-band observational
data. The maximum peak of the distribution $P(z)$ is marked as $z_s=\zbpz$ \citep{2012MNRAS.421.2355H}.

\subsection{Number-density Distribution of Source Galaxies}

For each cluster, we select a field-of-view (FOV) region around its center for weak lensing analyses.
As shown in Figure \ref{Fig:richz}, our lens clusters have a redshift distribution.
We therefore choose an FOV with a fixed physical area of $4\hmpc\times 4\hmpc$ for each cluster.
The angular region of the FOV then varies and is $(4\hmpc/D_d)^2$.
For paired lens systems, the FOV is centered at the midpoint of the two cluster centers, and the side angular length is $(R_{12}+4\hmpc)/D_d(z_1)$.

In each FOV, we only choose the selected CFHTLenS galaxies with photo $z$ in the range $(z_d+0.1,~1.3)$ as the background source galaxies,
where $z_d ~(>0.08)$ is the lens cluster redshift. For a paired lens system, $z_d$ is taken to be the highest redshift of the two.
The upper bound of this source redshift range is chosen according to the studies of \cite{2012MNRAS.427..146H} showing
that the photo-$z$ estimates in the range of $(0.2,~1.3)$ are reliable in comparison with the spectroscopic redshifts for CFHTLenS.
The lower bound $z_d+0.1$ is set to suppress the possible signal dilutions caused by foreground or member galaxies.

We estimate the effective number density of source galaxies in each FOV by

\begin{equation}\label{eq:numbg}
n_{\rm bg}=\frac{N_{\rm eff}}{(1-f_{\rm m})\Omega}
\end{equation}
where $\Omega$ and $f_{\rm m}$ are the total area and the mask fraction of the considered FOV, respectively.
The effective number of galaxies $N_{\rm eff}$ is estimated by, following \cite{2012MNRAS.427..146H},
\begin{equation}\label{eq:numeff}
N_{\rm eff}=\frac{(\sum\limits\omega_{i})^2}{\sum\omega_{i}^2}
\end{equation}
where $\omega_i$ is the lensfit weight for $i$th galaxy, and the sum is over all chosen background galaxies in the FOV.

\begin{figure}[!ht]
  \centering
  \includegraphics[width=0.45\textwidth]{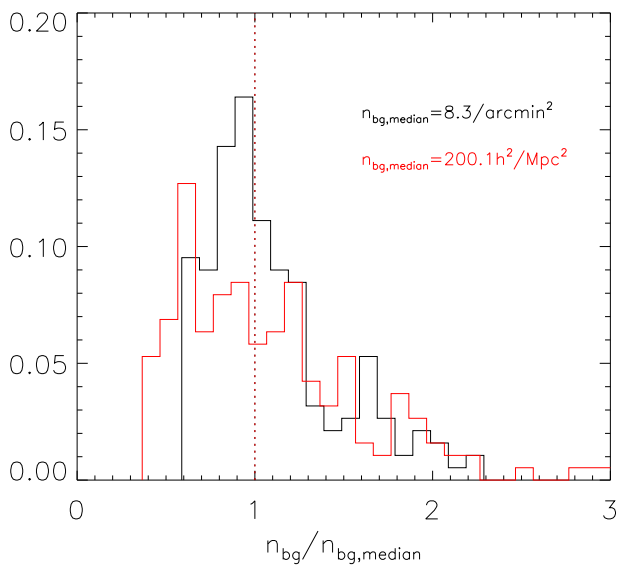}\\
  \caption{The normalized probability distribution of the effective number density $n_{\rm bg}$ scaled by the corresponding median value
$n_{\rm bg, median}$. Black and red histograms are respectively for the distribution
in angular size and physical size. The medians of the distributions are shown in the top right corner with the same color.
The vertical dotted lines indicate the position of medians of the distributions.}
\label{Fig:cfcsnbg}
\end{figure}

Figure \ref{Fig:cfcsnbg} shows the normalized probability distribution of $n_{\rm bg}$ for lens clusters.
The black distribution presents the distribution in units of $\amin^{-2}$ and the red is in units of $h^2\mpc^{-2}$.
The vertical dashed lines indicate the corresponding medians. We see that for the lens clusters in our consideration,
the median value of $n_{\rm bg}$ is $\sim 8.3 \hbox{ arcmin}^{-2}$, and about one-third of the clusters' FOVs
have $n_{\rm bg}>10 \hbox{ arcmin}^{-2}$.

As described above, our background galaxies for cluster weak lensing analyses are selected only based on their photometric redshifts.
Because of photo-$z$ errors, some foreground or member galaxies may be misidentified as background galaxies, which in turn can lead to
the dilution of lensing signals. To evaluate the possible contaminations, we look for concentrations of galaxies around the isolated clusters.

Specifically, we define a quantity $f_{\rm n}(R)$ as
\begin{equation}\label{eq:intelf}
f_{\rm n}(R)=\frac{n(R)}{n(R^\prime)}-1
\end{equation}
where $n(R)$ is the number density of targeted galaxies in a narrow annulus centered at radius $R$ from a cluster center,
and $R^\prime$ is adopted to be $2\hmpc$, the outer edge of our chosen FOV.

\begin{figure*}
  \centering
  \includegraphics[width=0.9\textwidth]{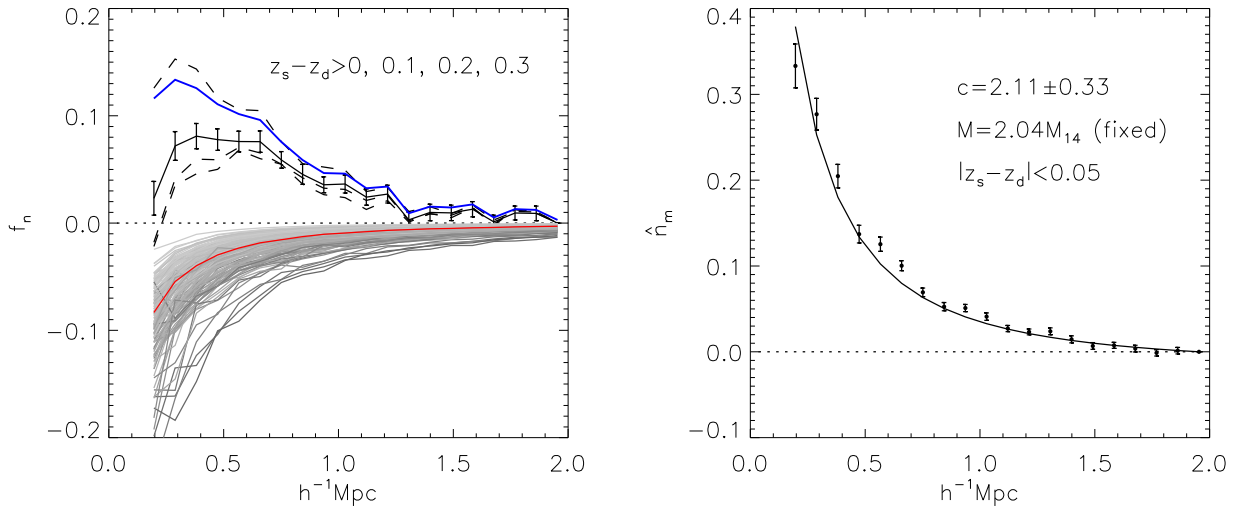}\\
  \caption{Left: the number density excess fraction $f_{\rm n}$ and the lensing magnification depletion fraction $f_{\mu}$.
The black lines show, from top to bottom, $f_{\rm n}$ for source galaxies with redshifts $z_{s,d}>0,~0.1,~0.2$, and $0.3$, respectively.
For clarity, we only show error bars for the case with $z_{s,d}>0.1$. The gray lines show the median depletion fraction $f_{\mu}$
for each of the $158$ clusters.
The red line presents the estimated median depletion fraction for the cluster sample. The blue line shows the corresponding median
interloper fraction $f_{\rm I}$ for $z_{s,d} > 0.1$.
Right: normalized number density $\hat{n}_m$ of galaxies with redshift in the range $|z_s-z_d|<0.05$. The solid line shows the NFW fitting by fixing the mass to be $2.04M_{14}$ (corresponding to richness $\lambda\sim34$).
  The errors in both panels are estimated from 100 bootstrappings. All of the clusters are stacked in this analysis.}\label{Fig:ngalr}
\end{figure*}

The black lines in the left panel of Figure \ref{Fig:ngalr} present the stacked $f_{\rm n}$ over all the
isolated lens clusters as a function of radius for source galaxies with
redshift cutoff $z_{s,d}=z_s-z_d>0,~0.1,~0.2$, and $0.3$ (from top to bottom),
respectively, where $z_s$ is taken to be $\zbpz$ from CFHTLenS.
For clarity, we only show error bars for the case with $z_{s,d}>0.1$, where the errors are estimated using 100 bootstrap
resamplings of the galaxies in the stack.
It is seen clearly that the higher the redshift cut for the source galaxies, the flatter the line is.
Note that we use the cut $z_{s,d}>0.1$ in our lensing analyses.
In this case, $f_{n}\sim 7\%$ in the inner part.
Our analyses for clusters in different richness bins show that $f_{\rm n}$ is nearly independent of
the cluster richness (i.e., mass).

It is noted that the definition of $f_{\rm n}$ in Equation (\ref{eq:intelf})
implicitly assumes a uniform distribution of background galaxies in the clusters' FOV. However, the lensing magnification effect of clusters can
result in a nonuniform distribution of background galaxies \citep[\eg][]{1995ApJ...438...49B,1997ARA&A..35..389E,
1999ARA&A..37..127M,2005MNRAS.361.1287M,2011ApJ...738...41U,2011ApJ...729..127U,2015arXiv150704493O,2015arXiv150704385U}.
The decrease of $f_{\rm n}$ in the very central region seen in Figure \ref{Fig:ngalr} is an indication of such an effect.
Therefore, to estimate the contamination fraction better, we need to take into account the nonuniform distribution of
background galaxies that is due to the lensing magnification bias \citep[\eg][]{2015arXiv150704493O}, which depends on the
mass and redshift of lens clusters and on the source redshift distributions.

Estimated from our Monte Carlo mock simulations to be detailed in Section 4,
the gray lines in the left panel of Figure \ref{Fig:ngalr} show the median depletion fraction $f_{\mu}$ of background galaxies
for each of the 158 isolated clusters, where $f_{\mu}=\mu(R)^{2.5s-1}-1$, $\mu=1/[(1-\kappa)^2-|\gamma|^2]$ is the lensing magnification,
and $s$ depends on the luminosity function of background galaxies and is taken to be $s=0.15$ in our analyses
\citep[\eg][]{1995AIPC..336..320B,2014ApJ...795..163U,2015arXiv150704385U}.
Darker lines are for clusters with higher richness.
It is seen that more massive clusters generate a larger magnification bias.
The red line shows the median value of $f_{\mu}$ of all of the $158$ clusters with $f_{\mu}\sim -8\%$ in the innermost bin.
This is much smaller than that shown in \citet{2015arXiv150704493O} for their
X-ray-luminous clusters. Our redMaPPer cluster sample has a median richness of $\lambda \sim 34$, which corresponds
to the mass of $M_{200}\sim 2\times 10^{14}h^{-1}M_{\odot}$ according to the mass--richness relation shown in Section 3.2.1,
in contrast to $M_{200}\sim 7\times 10^{14}h^{-1}M_{\odot}$ for the sample used in \citet{2015arXiv150704493O}.
Furthermore, the CFHTLenS weak lensing observations used in our analyses are shallower than their targeted observations
with Suprime-Cam on Subaru. The smaller lens clusters and the lower source redshifts result in a lower level of
magnification bias.

Including the lensing magnification depletion effect, the contamination fraction can then be estimated by
\begin{equation}\label{eq:mbias}
f_{\rm I}(R)=\frac{1+f_{\rm n}(R)}{1+f_\mu(R)}-1.
\end{equation}
The blue line in Figure \ref{Fig:ngalr} shows median $f_{\rm I}$ for $z_{s,d}>0.1$ by using $f_{\rm n}$ of the black solid line and
$f_{\mu}$ of the red line. We will discuss the dilution effect with the contamination rate estimated without and with
the magnification bias included, respectively, in Section 4.

As a further test, we also analyze the normalized surface number density profile for galaxies lying within the range $-0.05<z_{s,d}<0.05$. It is defined as
\begin{equation}\label{eq:nmemr}
\hat{n}_m(R)=\frac{n_m(R)-n_m(R^\prime)}{N_{m,{\rm tot}}-\pi {R^\prime}^2n_m(R^\prime)}
\end{equation}
where $n_m(R)$ is the number density in an annulus of radius $R$, $N_{m,{\rm tot}}$ denotes the total number of targeted galaxies within radius $R^\prime$, which is fixed to be $2\hmpc$. The stacked $\hat{n}_m(R)$ over our lens clusters is shown in the right panel of Figure \ref{Fig:ngalr}.
This distribution can be related to the surface number density distribution of cluster member galaxies.

The number-density profile of member galaxies for clusters of galaxies has been studied extensively based on observations and simulations
\citep[\eg][]{2004ApJ...610..745L,2005MNRAS.356.1233V,2006ApJ...647...86C,2012MNRAS.423..104B,2012MNRAS.427..428G,2014Natur.509..177V}.
It is also popularly modeled as an NFW profile. In accord with the definition of $\hat{n}_m(R)$,
we normalize the projected NFW density profile to define $\hat{\Sigma}(R)$ as follows:
\begin{equation}\label{eq:nsigr}
\hat{\Sigma}(R)=\frac{\Sigma_{\rm NFW}(R)-\Sigma_{\rm NFW}(R^\prime)}{\Sigma_{\rm tot}-\pi {R^\prime}^2\Sigma_{\rm NFW}(R^\prime)}
\end{equation}
where $\Sigma_{\rm NFW}(R)$ is the NFW surface number density at radius $R$ and $\Sigma_{\rm tot}=2\pi\int_0^{R^\prime}R\Sigma_{\rm NFW}(R)dR$.
When fitting $\hat{\Sigma}(R)$ to the stacked $\hat{n}_m(R)$, we obtain an estimate for the average concentration parameter
to be $c\approx 2.11\pm0.33$. Note that here we fix the mass to be $M=2.04M_{14}$ ($M_{14}\equiv10^{14}h^{-1}\msun$), corresponding to
the median richness of the sample $\lambda\sim34$ from the richness--mass relation
under the cosmology we consider. We will see in Section 5 that this result is consistent with our weak lensing analyses.

We note that to obtain the stacked $f_{\rm n}$ and $\hat{n}_m(R)$ shown in the left and right panels of Figure \ref{Fig:ngalr},
we assume that the optical center for each cluster is the true center without considering the possible off-center effect.
Its impact on our weak lensing analyses will be analyzed in Section 5. The off-center effect can also
change slightly the profile of $f_{\rm n}$. Because this effect is much smaller than
the large statistical uncertainties of $f_{\rm n}$ for our cluster sample,
we neglect the off-center effect on $f_{\rm n}$ in our studies here.

\section{Weak Lensing Analysis}

In this section, we describe our weak lensing analyses for the selected clusters.
After a brief introduction of the weak lensing theory, we present details of the NFW fitting method.

\subsection{Weak Lensing Effects from Lens Clusters}

Massive clusters are the dominant contributors to their LOS weak lensing signals.
For each cluster, its lensing effect on a background source can be described by using the thin lens approximation, in which
the lensing potential is given by \citep{2001PhR...340..291B}
\begin{equation}
\boldsymbol \nabla^2\Phi=2\kappa,
\label{eq:lenspot}
\end{equation}
where $\kappa=\Sigma/\Sigma_{\rm crit}$ is the lensing convergence, with $\Sigma$ being the LOS projected surface density of the cluster.
The critical surface density is defined by
\begin{equation}\label{eq:criden}
\Sigma_{\rm crit}=\frac{c^2}{4\pi G}\frac{D_s}{D_dD_{ds}}
\end{equation}
where $D_s$, $D_d$, and $D_{ds}$ are the angular diameter distances from the observer to the source, to the lens, and
from the lens to the source, respectively. The induced observational effects can be described by the Jacobian matrix
\begin{eqnarray}\label{eq:jac}
\nonumber  \mathcal{A}=\left(
              \begin{array}{cc}
                1-\kappa-\gamma_1 & -\gamma_2 \\
                -\gamma_2 & 1-\kappa+\gamma_1 \\
              \end{array}
            \right) \\
  ~=(1-\kappa)\left(
              \begin{array}{cc}
                1-g_1 & -g_2 \\
                -g_2 & 1+g_1 \\
              \end{array}
            \right)
\end{eqnarray}
where $\gamma_i$ $(i=1,2)$ are the lensing shear components with $\gamma_1=(\partial^2_{11}\Phi-\partial^2_{22}\Phi)/2$ and
$\gamma_2=\partial^2_{12}\Phi$, and $g_i=\gamma_i/(1-\kappa)$ is the corresponding reduced shear component.

The lensing effect from a lens cluster leads to a shape distortion for a background galaxy.
Its observed ellipticity is then given by
\begin{equation}\label{eq:ellip}
\boldsymbol e=
\displaystyle\frac{\boldsymbol e_s+\boldsymbol g}{1+\boldsymbol g^\ast\boldsymbol e_s},
\end{equation}
where the ellipticity $\boldsymbol e$ is defined by
\begin{equation}
\boldsymbol e =e_1+ie_2= \frac{1-b/a}{1+b/a}\exp(2i\varphi),
\end{equation}
with $b/a$ and $\varphi$ being the axial ratio and the orientation angle of the approximate ellipse of the observed image derived
from the quadrupole moments of its light distribution. The intrinsic ellipticity $\boldsymbol e_s$ is defined in the same way
with respect to the unlensed image.

It has been shown that without considering the intrinsic alignments of source galaxies, the expectation value of the
above defined ellipticity is an unbiased estimate of $\boldsymbol g$ \citep{1997A&A...318..687S}. Therefore we can derive
lensing signals from the observed $\boldsymbol e$ and further constrain the density profile of a lens cluster.

As shown in Section 2, there are paired clusters in our lens sample. For these systems, both clusters contribute to the lensing signals.
In the weak lensing limit, the distortions caused by the two lenses are additive  \citep{2001MNRAS.326.1015C,2012A&A...546A..79I}:
\begin{equation}\label{eq:resh2}
\boldsymbol g \simeq \boldsymbol g^{(1)}+\boldsymbol g^{(2)}
\end{equation}
where $\boldsymbol g^{(1)}$ and $\boldsymbol g^{(2)}$ are the reduced shears from the two clusters at redshift $z_1$ and $z_2$, respectively.
Equation (\ref{eq:resh2}) will be used in analyzing paired clusters.

Observationally, source galaxies span a range of redshifts. Because of the redshift dependence of the lensing signals
as seen from Equation (\ref{eq:criden}), galaxies at different redshifts experience different lensing effects even if the lens is the same.
To separate the redshift dependence of the signals from their dependence on the lens density profile,
a scale factor can be introduced in $\boldsymbol g$:

\begin{equation}\label{eq:resh1}
\boldsymbol g=\frac{\boldsymbol\gamma}{1-\kappa}=\frac{F\boldsymbol\gamma_f}{1-F\kappa_f}
\end{equation}
where $F=\Sigma_f\Sigma_{crit}^{-1}$. In this paper, we fix $\Sigma_f$ to be $1 \times 10^{15}\msun\mpc^{-2}$.
The scaled quantities $\gamma_f$ and $\kappa_f$ then depend only on the density profile of the lens.
According to pad\'{e} approximation \citep{1992nrfa.book.....P,1997A&A...318..687S}, the redshift average of reduced shear
$\boldsymbol g$ in a local region can be written as
\begin{equation}\label{eq:mresh}
\langle\boldsymbol g\rangle=\frac{\langle F\rangle\boldsymbol\gamma_f}{1-\kappa_f\langle F^2\rangle/\langle F\rangle}
\end{equation}
where angle brackets $\langle~\rangle$ denote the average of the quantities over the redshift distribution of source galaxies.
Equation (\ref{eq:mresh}) will be used to model the lensing signals theoretically, to be compared with observational data.

\subsection{NFW Fitting}

As seen from Equation (\ref{eq:mresh}), in order to model the lensing signals expected from a lens cluster, we need
the quantities $\langle F\rangle$ and $\langle F^2\rangle$. We estimate them from the CFHTLenS galaxies used in
our weak lensing analyses. Specifically, for a source galaxy in CFHTLenS, a photo-z distribution $P(z)$ is given in addition to
the best-fit $\zbpz$ \citep{2012MNRAS.421.2355H}.  To utilize this information in the analyses,
we calculate the critical surface density for each source galaxy by
\begin{equation}\label{eq:criden1}
\bar \Sigma_{\rm crit}^{-1}=\int_{z_d}^\infty dz\Sigma_{\rm crit}^{-1}(z_d,z)P(z)
\end{equation}
where $z_d$ is the redshift of the considered lens cluster. The corresponding scale factor is $F=\Sigma_f\bar\Sigma_{\rm crit}^{-1}$.

The averages of $F$ and $F^2$ are then estimated by
\begin{equation}\label{eq:means1}
\langle F\rangle=\frac{\sum\omega_iF_i}{\sum\omega_i},\,\,\langle F^2\rangle=\frac{\sum\omega_iF_i^2}{\sum\omega_i},
\end{equation}
where the sum is over all the source galaxies in each of the radial bins or grids for isolated clusters or paired clusters, respectively,
and $\omega_i$ is the weight of the $i$th source galaxy from lensfit.

The observed lensing signals are estimated from the galaxy ellipticities given in the CFHTLenS data catalog.
We pay attention to the corrections of the additive bias $c_2$ and the multiplicative bias $m$ in the shear measurements.
Following \cite{2013MNRAS.429.2858M}, we correct the additive bias $c_2$ to the $e_2$ component for each source galaxy
(hereafter we denote this corrected one as $e_2$). The correction for the multiplicative bias should be done statistically.
The average of the observed $\boldsymbol e$ is then calculated by
\begin{equation}\label{eq:means2}
\langle \boldsymbol e\rangle=\frac{\sum\omega_i\boldsymbol e_i}{\sum\omega_i(1+m_i)}
\end{equation}
where again the sum is over the source galaxies in each of the radial bins or grids.

For an isolated lens, we analyze the average tangential shear signals in different radial bins around the lens center.
Therefore the tangential component $e_t$ instead of $\boldsymbol e$ is used in Equation \ref{eq:means2}.
The tangential component $e_t$ for each galaxy is calculated by
\begin{equation}\label{eq:tansh}
e_t=-[e_1{\rm cos}(2\phi)+e_2{\rm sin}(2\phi)],
\end{equation}
where $\phi$ is the angle of the line connecting the source galaxy and the lens center.

\subsubsection{NFW Fitting for Isolated Clusters}

Our studies aim to constrain $(M, c)$ for each lens cluster from weak lensing analyses and subsequently to derive the $M$--$c$ relation
for the lens sample.

In \citet{2014ApJ...785...57D}, we show that the shape noise leads to strongly correlated errors in the weak-lensing-constrained $(M,c)$
for individual clusters due to the degenerate dependence of the reduced shear on the two parameters $M$ and $c$.
This in turn can result in a significant bias in the derived $M$--$c$ relation. The higher the shape noise is, the larger the bias, with notably
a steeper slope for the fitted power-law relation of $c(M)$ in comparison with the underlying true $M$--$c$ relation.
To account for the noise effect, we introduced a Bayesian method that includes the correlated error matrix
in comparing a model $(M,c)$ to the weak-lensing-constrained $(M,c)$.  In that, we assumed that the error distribution of each individually
constrained $(M,c)$ follows the two-dimensional (2D) Gaussian distribution in log space. We showed that the approximation works well
for cases with $n_{\rm bg}>10~\amin^{-2}$. We also pointed out that including cluster observables that are good mass indicators in
weak lensing analyses as priors can help to suppress the shape noise effect significantly \citep{2014ApJ...785...57D}.

In our current studies with CFHTLenS, as shown in Figure \ref{Fig:cfcsnbg}, the typical number density of source galaxies used
in the analyses is $n_{\rm bg}\sim 8\hbox{ arcmin}^{-2}$, and the resulting shape noise is rather high.
Therefore its bias effect on the derived $M$--$c$ relation is significant as expected. Because of the high noise level, the Gaussian approximation
about the error distribution of $(M,c)$ may not be valid. Thus here we will not use the Bayesian method developed in \citet{2014ApJ...785...57D}
to correct for the bias effect from the shape noise. Instead, when constraining $(M,c)$ for each lens cluster from weak lensing data,
we will employ a prior on $M$ based on the mass--richness ($M$--$\lambda$) relation of redMaPPer clusters.

\citet{2015MNRAS.449.1897O} carefully examined different mass estimators based on cluster observables using simulated clusters.
They concluded that the richness-based mass estimate methods perform the best. Particularly for the richness determined by the redMaPPer algorithm,
they analyzed the mass determination using the abundance-matching method. It is shown that
no systematic bias exists for the estimated mass \citep{2015MNRAS.449.1897O}.
Therefore here we adopt the abundance-matching method under the cosmological model we considered to derive the $M$--$\lambda$ relation for redMaPPer clusters.
Specifically, we use clusters with $z<0.35$, which are shown to be complete for the catalog \citep{2014ApJ...785..104R}, to estimate the $M$--$\lambda$ relation.
We follow the procedures described in the Appendix B of \citet{2012ApJ...746..178R}. The intrinsic scatter of mass given a richness
is taken to be $\sigma_{M|\lambda}\simeq0.25$, and the mass function of \citet{2008ApJ...688..709T} is adopted.
The obtained $M$--$\lambda$ relation is
\begin{equation}\label{eq:mrch}
\ln(M_\lambda/\hms)\simeq28.89+1.15\ln(\lambda)
\end{equation}
where $M_\lambda$ is in accord with $M_{200}$. In the following analyses, we assume that there is no evolution for this $M$--$\lambda$ relation
in the redshift range of the clusters in $z=[0.08, 0.55]$, although the evolution may exist \citep[\eg][]{2014MNRAS.444..147O}.

Thus the mass prior employed in our analyses is given by
\begin{equation}\label{eq:probmc2}
P(M|\lambda)=\frac{1}{\sqrt{2\pi}\sigma_{M|\lambda}}\exp\left[-\frac{(\ln M-\ln M_\lambda)^2}{2\sigma_{M|\lambda}^2}\right]
\end{equation}
where $\sigma_{M|\lambda}=0.33$ is adopted for fitting so as to account for possible systematic uncertainties \citep{2012ApJ...746..178R}.

Besides the shape noise, the center misidentification can also lead to a significant effect on the weak-lensing-derived $(M,c)$ for a lens cluster.
In our study here, we choose the position of the galaxy with the highest centering probability $P_{\rm cen}$ as the center of the cluster mass distribution.
Analyses have shown, however, that such optical centers can deviate from the true mass centers of clusters \citep[\eg][]{2007arXiv0709.1159J,2010MNRAS.405.2215O}.
To include the off-centering effect, as a prior we model the offset probability distribution for the optical center of each lens cluster to be

\begin{align}\label{eq:proff}
P(R_{\rm off})& =P_{\rm cen}P(R_{off}|0.09\hmpc)+ \nonumber\\[10pt]
  {} & (1-P_{\rm cen})P(R_{off}|0.42\hmpc)
\end{align}
where $P_{\rm cen}$ is the redMaPPer centering probability of the chosen central galaxy, and $1-P_{\rm cen}$ indicates the probability
that this galaxy is misidentified as the center.
The probability function $P(R_{off}|\sigma_s)$ follows the 2D Gaussian form given by
\begin{equation}\label{eq:proff1}
P(R_{off}|\sigma_s)=\frac{R_{off}}{\sigma_s^2}\exp\left[-0.5(R_{off}/\sigma_s)^2\right]
\end{equation}
where $\sigma_s=0.09\hmpc$ is adopted for true optical centers and $\sigma_s=0.42\hmpc$ for misidentified centers \citep{2010MNRAS.405.2215O}.
We note that the information of $P_{\rm cen}$ given to each central galaxy candidate in the redMaPPer algorithm is crucial.
Only with this information are we able to include the off-centering probability in individual cluster studies.
For the redMaPPer cluster catalog we used, the median value of $P_{\rm cen}$ is $\sim0.92$.

Based on the above priors, we can then model the probability for deriving the underlying $(M,c,R_{\rm off})$
given the observed tangential shear signals $\langle e_t\rangle$ and the cluster richness $\lambda$ to be
\begin{equation}\label{eq:probmc}
P(M,c,R_{\rm off}|\langle e_t\rangle,\lambda)\propto P(\langle e_t\rangle|M,c,R_{\rm off})P(M|\lambda)P(R_{\rm off})
\end{equation}
where
\begin{align}\label{eq:probmc1}
P(\langle e_t\rangle|M,c,R_{\rm off}) \hspace{5cm} & {} \nonumber\\[10pt]
=\prod_i \frac{1}{\sqrt{2\pi}\sigma_i}\exp\left\{-\frac{[\langle e_t\rangle_i-g_{t,i}(M,c,R_{\rm off})]^2}{2\sigma_i^2}\right\} & {}
\end{align}
where the index $i$ denotes the $i$th radial bin, and the error in each bin is calculated by $\sigma_i=\sigma_{\boldsymbol e}/\sqrt{2N_{{\rm eff},i}}$
with $N_{{\rm eff},i}$ being the effective number of galaxies in the $i$th radial bin. We take $\sigma_{\boldsymbol e}$ to be $0.4$.
The theoretical value $g_{t,i}$ is the averaged tangential reduced shear in the $i$th radial bin $R_i$ relative to the offset center indicated by
$R_{\rm off}$ using the NFW model. It is given by

\begin{equation}\label{eq:gtoff}
g_{t,i}=\frac{1}{N}\sum\limits_{j=0}^{N}\frac{\gamma_{f,t}(R_i,\theta_j|R_{\rm off})\langle F\rangle_i}{1-\kappa_f(R_i,\theta_j|R_{\rm off})\langle F^2\rangle_i/\langle F\rangle_i}
\end{equation}
where $\gamma_{f,t}(R_i,\theta_j|R_{\rm off})$ and $\kappa_f(R_i,\theta_j|R_{\rm off})$ are respectively the tangential shear and convergence
at $(R_i,\theta_j)$ relative to the chosen center. Note that we allow the chosen center to be offset from the true center.
In this case, the theoretical reduced shear signals around the wrong center are not axisymmetric anymore, and
therefore we need to write explicitly the dependence on the azimuthal coordinate $\theta_j$. Here we divide the
azimuthal direction into $N$ parts, and $\theta_j=2\pi j/N$. Although different values of $N$ give rise to slightly different results, especially
for the cases with large center offsets, the differences are not significant. In the following analyses, we take $N=100$.
For the radial bins, we choose galaxies in the radial range from $0.15$ to $2\hmpc$ and then divide them into 10 equal radial bins.

Based on above equations, we can then estimate the parameter values $M$, $c$, and $R_{\rm off}$ for each lens cluster
by maximizing the likelihood function $\ln P(M,c,R_{\rm off}|\langle e_t\rangle,\lambda)$ using Markov chain Monte Carlo (MCMC) algorithm.
In the calculations, we use a uniform prior for $c$ in the range of $(0.01,~99)$.

If we assume the optical centers are located at the true halo centers, \ie~ without considering the off-centering effect,
Equation \ref{eq:probmc} and \ref{eq:gtoff} will be reduced to
\begin{equation}
P(M,c|\langle e_t\rangle,\lambda)\propto P(\langle e_t\rangle|M,c)P(M|\lambda)
\end{equation}
and
\begin{equation}
g_{t,i}=\frac{\gamma_{f,t}(R_i)\langle F\rangle_i}{1-\kappa_f(R_i)\langle F^2\rangle_i/\langle F\rangle_i}
\end{equation}

We also consider the simple fitting without using the mass priors from the mass--richness relation.
For this, we choose a uniform prior for mass in the range of $(10^{11},~10^{16})\hms$.
This is the traditional method often used in the fitting.
We name this fitting without considering the center offset and the mass prior as the direct fitting method.

\begin{figure*}
  \centering
  \includegraphics[width=0.8\textwidth]{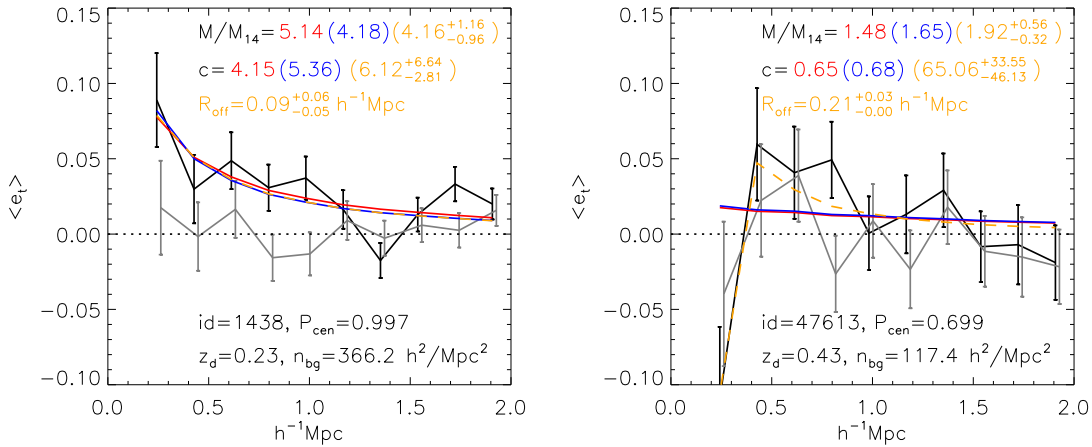}\\
  \caption{NFW fitting for isolated clusters from CFHTLenS. Black data points with error bars are the observed mean tangential shears in each radial bin. Gray lines are for the rotated component of reduced shear. The red, blue, and orange lines show the fitting result with different priors. The red is for direct fitting. Blue shows the result with prior $P(M|\lambda)$ on mass. The orange line indicates the result with $P(R_{\rm off})P(M|\lambda)$. The corresponding results are plotted at the top of the panel with the same colors as the lines. At the bottom in each panel, we also show the cluster redMaPPer id, redshift $z_d$, background number density $n_{\rm bg}$, and the centering probability $P_{\rm cen}$. In the innermost bin, the effective number of background galaxies is $\sim82$ for cluster $1438$ and $\sim35$ for $47613$.}\label{Fig:cfhtgtr}
\end{figure*}

Figure \ref{Fig:cfhtgtr} shows two examples of isolated clusters from CFHTLenS, one at $z_d= 0.23$ (left) and one at $z_d=0.43$ (right).
The black data points with error bars are the observed mean tangential shears in each radial bin. The gray lines present the trend of
a $45^\circ$-rotated component of the reduced shear for comparison.
The rotated component is consistent with zero within 1$\sigma$ error, indicating a negligible systematic effect on the
shear signal measurements. The red, blue, and orange lines show the fitting results with different priors. The red is for the direct fitting.
Blue shows the result with a prior $P(M|\lambda)$ on mass. The orange line is the result including both the
center-offset prior and the mass prior with $P(R_{\rm off})P(M|\lambda)$. The corresponding fitted values are written
at the top of each panel with the same colors as the lines. At the bottom of each panel, we show the cluster's redMaPPer id, redshift $z_d$,
background number density $n_{\rm bg}$ and the centering probability $P_{\rm cen}$. It is seen that the concentration tends to be higher
by taking into account the off-centering problem, especially for cases with relatively low $P_{\rm cen}$.
For the cluster shown in the right panel, it is at a relatively high redshift $z_d=0.43$. The number density of source galaxies is rather low,
and the centering probability is also low. Thus the results are affected severely by the shape noise and the center offset.

\subsubsection{NFW fitting for paired clusters}

For paired clusters, we simultaneously fit the profiles of the two clusters using the two-dimensional lensing signals.
With still noisy data, we are not allowed to consider too many free parameters.
Therefore, for paired clusters, we do not take into account the center-offset problem and assume that the two optical centers are the true centers
of the corresponding clusters.

For a paired lens cluster system, we divide its FOV into $30\times30$ regular grids and obtain the average of galaxy ellipticities
$\langle\boldsymbol e\rangle$ in each grid using background galaxies therein. The corresponding probability for the $(M_1,c_1,M_2,c_2)$ estimate given
the observed $\langle \boldsymbol e\rangle$, and the richness $\lambda_1$ and $\lambda_2$ is then
\begin{align}\label{eq:probmc2d}
P(M_1,c_1,M_2,c_2|\langle \boldsymbol e\rangle,~\lambda_1, ~\lambda_2)\hspace{3cm} & {} \nonumber\\[10pt]
\propto P(\langle \boldsymbol e\rangle|M_1,c_1,M_2,c_2)P(M_1|\lambda_1)P(M_2|\lambda_2) & {}
\end{align}
where the subscripts ``1'' and ``2'' mark the quantities for clusters at $z_1$ and $z_2$ respectively.
The probability of the observed $\langle \boldsymbol e\rangle$ is written as
\begin{equation}\label{eq:probmc2d1}
P(\langle \boldsymbol e\rangle|M_1,c_1,M_2,c_2)=
\prod_i \frac{1}{\sqrt{2\pi}\sigma_i}\exp\left\{-\frac{|\langle \boldsymbol e\rangle_i-\boldsymbol g_i|^2}{2\sigma_i^2}\right\}
\end{equation}
where $i$ indicates the $i$th grid, and $\boldsymbol g_i=\boldsymbol g^{(1)}_{i}(M_1,c_1)+\boldsymbol g^{(2)}_{i}(M_2,c_2)$ is the summation of
theoretical reduced shears from the two clusters. We adopt $\sigma_i=0.4/\sqrt{N_{{\rm eff},i}}$ for each grid.
In the fitting, we exclude the innermost five grids around the center of each cluster.

\begin{figure*}
  \centering
  \includegraphics[width=0.8\textwidth]{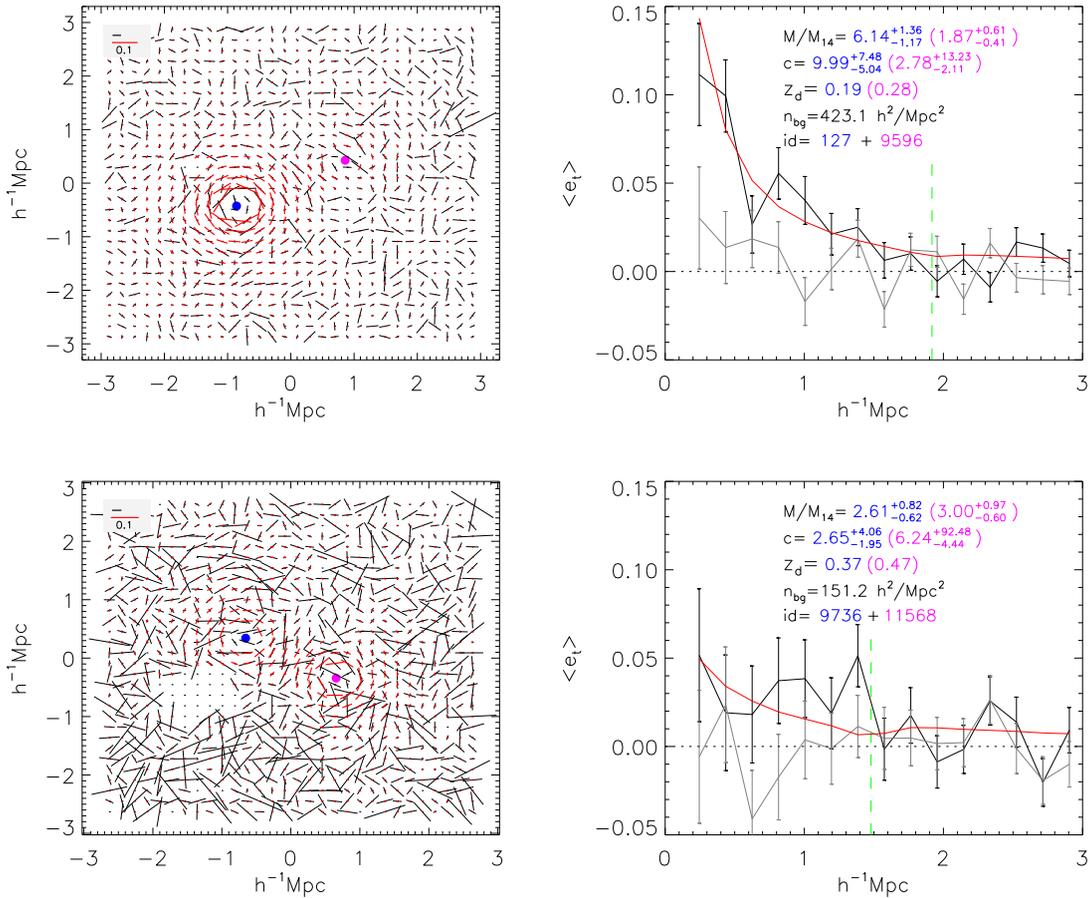}\\
  \caption{NFW fitting with prior $P(M|\lambda)$ for paired clusters in CFHTLenS. The top and bottom panels are for pairs $127+9596$ and $9736+11568$, respectively. Left panels show the two-dimensional shear field for each pair. Black sticks display the observed shear field. The direction of a stick indicates the direction of mean shape of galaxies in the grid. The estimated shear field is plotted with red sticks. The grids without sticks are due to the masked regions where there are no galaxies. The black and red sticks in the top left corner respectively indicate the scale value of $0.1$ for corresponding field. The blue and magenta filled circles mark the optical centers of clusters. Blue is for low-redshift clusters and magenta for the higher redshift cluster. The estimated mass and concentration are presented in the right panel with consistent colors. The right panels show the tangential shears (black data points) relative to the center of low-redshift cluster. Gray data points are for the rotated shear component. The red line is the predicted trend with parameter values estimated from the 2D field. The vertical dashed green line points out the projected distance between the two clusters in the low-redshift lens plane. The cluster id, redshift $z_d$, and number density $n_{\rm bg}$ are also shown in the right panels.}\label{Fig:cfhtgg}
\end{figure*}

Figure \ref{Fig:cfhtgg} presents two examples of paired lens systems.
The left panels show the two-dimensional ellipticities. Black sticks display the observed $\langle \boldsymbol e\rangle$ field and
the red ones for the reduced shear field of $\langle \boldsymbol g\rangle$ from fitting.
The grids without black sticks are due to the masked regions where there are no source galaxies. The black and red horizontal lines
in the top left corner respectively indicate the scale value of $0.1$ for the corresponding field. The blue and magenta filled circles mark
the optical centers of the clusters. The blue one is for the low-redshift cluster and the magenta one is for the high-redshift counterpart.
The estimated mass and concentration for the paired clusters are presented in the right panel with the same colors.
In the right panels, we also show the tangential reduced shear signals (black data points) relative to the center of the low-redshift cluster.
The gray data points are for the $45^\circ$-rotated component. The red line is the result using the parameters estimated from fitting to the 2D shear field.
The vertical dashed green line marks the projected distance between the two clusters in the low-redshift lens plane.

The centering probabilities $P_{\rm cen}$ for clusters in the pair $127+9596$ (upper panels) are respectively $\sim0.85$ and $\sim0.80$.
For the pair $9736+11568$ (lower panels),  they are $P_{\rm cen}\sim0.98$ and $\sim0.95$, respectively.
For the pair $9736+11568$, there is a relatively large masked region in the field.
We will see from the Monte Carlo simulation analyses, shown in the next section,
that the existence of masks away from the central region (note that we exclude the clusters with large central masked regions in
our lens sample) do not introduce a significant bias in the derived $M$--$c$ relation.

\section{Monte Carlo Simulations}

In order to test our fitting method and demonstrate the significance of properly taking into account the effects of shape noise and center offset,
we generate Monte Carlo simulations to mimic the observations, including photo-$z$ errors. The steps of generating mock catalogs are as follows:
\begin{enumerate}
  \item For each considered redMaPPer cluster, according to its richness $\lambda$, we randomly assign it a mass $M_\smT$
through Equations \ref{eq:mrch} and \ref{eq:probmc2} with the intrinsic scatter $\sigma_{M|\lambda}=0.25$.
  \item We assume a spherical NFW profile for each simulated cluster. Its concentration parameter $c_\smT$ is given randomly according to a log-normal
probability distribution with an intrinsic scatter of $\sigma(\log \ct)=0.12$. The mean value satisfies the following $M$--$c$ relation:
\begin{equation}\label{eq:cmp}
\lgg\,\ct=\lgg\,A+\alpha \lgg\,\frac{\mt}{M_{\rm p}}
\end{equation}
where $A=A_0/(1+z_d)^{B}$. We adopt $A_0=4$, $B=0.5$, $\alpha=-0.1$, and $M_p=M_{14}=10^{14}\hms$ as our fiducial input.
This power-law form is taken based on the analyses of \citet{2008MNRAS.390L..64D}. We approximately adjust it to the one in the {\it Planck} cosmology.
Specifically, recent simulations show that the normalization of the $M$--$c$ relation in {\it Planck} cosmology is $10\%-20\%$ higher than
that in WMAP5 cosmology for massive halos \citep{2014MNRAS.441.3359D,2014MNRAS.441..378L,2015ApJ...799..108D}.
We therefore multiply the normalization of Duffy's $M$--$c$ relation for full samples by $115\%$ to match approximately
the amplitude in {\it Planck} cosmology. We then obtain the normalization $A_{3D}\sim4.5$ relative to pivot mass $M_p=M_{14}$.
Note that we adopt perfectly spherical NFW halos in generating mocks here. In this ideal case,
the 2D weak-lensing-derived $M$--$c$ relation should be the same as the $M$--$c$ relation of 3D halos. However,
real halos have complicated mass distributions. It has been shown that the nonsphericity of halos, the existence of substructures,
and the effect of surrounding large-scale structures can lead to a negative bias in the 2D weak-lensing-derived $M$--$c$ relation
in comparison with that of 3D halos \citep{2007MNRAS.380..149C,2012MNRAS.421.1073B,2012MNRAS.426.1558G,2014ApJ...785...57D}.
Because our halo mocks do not include the complexity of their mass distributions, this negative bias will not
occur in our weak-lensing-derived $M$--$c$ relation for simulated clusters. In order to compare with observational results
more consistently, where this negative bias is expected, we further introduce a $10\%$ negative bias to the
normalization of the above $M$--$c$ relation. These considerations lead to $A_0=4$, used in our Monte Carlo simulations here.
  \item In the lens plane of each simulated cluster, we choose its optical center to be the same as that of its real redMaPPer counterpart.
For the true center, we consider two cases to demonstrate the off-center effects.
One is to assume that the optical center is the true center. The other is to randomly assign a true central position $(x_c,y_c)$
around its optical center for the simulated cluster  according to Equation \ref{eq:proff} with the centering probability $P_{\rm cen}$
from the corresponding redMaPPer cluster.
  \item In the FOV of each simulated cluster, the mock source galaxies are populated in accord with the
the CFHTLenS galaxies with the same angular positions and weights.
For the redshift of a mock galaxy $z_{s,{\rm sim}}$, it is given randomly based on the photo-$z$ distribution $P(z)$ of the corresponding CFHTLenS galaxy.
If $z_{s,{\rm sim}}<z_d$, there are no lensing signals for the mock galaxy. For $z_{s,{\rm sim}}>z_d$, the lensing signals from the simulated cluster are calculated.
We then assign an ``observed'' ellipticity to each of the mock galaxies according to Equation \ref{eq:ellip}.
The intrinsic ellipticity of source galaxies follows the probability density distribution
\begin{equation}\label{eq:pes}
    P(|\boldsymbol e_s|)=2\pi|\boldsymbol e_s|\frac{\exp\left(-|\boldsymbol e_s|^2/\sigma_{\boldsymbol e_s}^2\right)}
    {\pi\sigma_{\boldsymbol e_s}^2\left[1-\exp(-1/\sigma_{\boldsymbol e_s}^2)\right]},
\end{equation}
where $\sigma_{\boldsymbol e_s}$ is the rms of the total $\boldsymbol \epsilon_s$, which is taken to be $\sigma_{\boldsymbol e_s}=0.4$ in the default case.
For comparison, we also consider the case without intrinsic ellipticities. For that we set $\boldsymbol e_s=0$.
For mock galaxies, no multiplicative and additive biases are introduced.
It should be noted that we assign a decisive redshift $z_{s,{\rm sim}}$ to a mock galaxy for the purpose
of calculating the lensing signals on it. In the subsequent mock weak lensing analyses, we still
use the observed redshift $z_s$ and the probability $P(z)$ of the corresponding CFHTLenS galaxies to mimic
the real observational studies.
\end{enumerate}

To fully explore the effects of shape noise and center offset, following the above procedures, we generate four sets of mocks.
The fiducial one contains both the shape noise and the center-offset effects. We also generate a set of mock simulations
without shape noise but with the center-offset effect included. The other two sets assume no center-offset effect: we
choose the optical center of a cluster to be the true center of the corresponding simulated cluster, and with
and without shape noise. Each set of mock simulations has 100 realizations with respect to the lens cluster mass and
center assignments and the source galaxy intrinsic ellipticity and redshift assignment. Each realization contains 158 isolated mock clusters
and 31 paired ones as our real cluster catalog.

We note that we choose source galaxies for a cluster by setting a redshift lower cut of $z_s>z_d+0.1$.
For CFHTLenS photo-$z$ measurements, the scatter is $\sim 0.05$ \citep{2012MNRAS.421.2355H}. Therefore we expect the cut
to be able to suppress the dilution effect from member or foreground galaxies. However, as shown in Figure \ref{Fig:ngalr},
there is still a weak concentration for the selected source galaxies around the lens clusters with the number density excess $f_n\sim 0.07$.
Although the statistical fluctuations of $f_{\rm n}$ are large from cluster to cluster,
no significant richness dependence of $f_{\rm n}$ is detected.
The existence of the lensing magnification bias tends to decrease the observed number of background galaxies toward
cluster central regions, leading to higher contamination fractions $f_{\rm I}$ than $f_{\rm n}$ as shown in Equation (\ref{eq:mbias}).
In order to evaluate the dilution effect, we generate two other sets of mock simulations
for isolated clusters as follows.

We first estimate the lensing magnification of redMaPPer clusters for CFHTLenS observations from our mock clusters.
As described above, each mock cluster is given a redshift and a mass in accord with the redshift and the richness of the
real redMaPPer cluster in the FOV. The angular positions and redshifts of mock source galaxies are also assigned based on the
angular positions and redshift distribution of CFHTLenS galaxies.
Therefore for each source galaxy, we can calculate its lensing magnification $\mu$,
and furthermore the depletion fraction $f_{\mu}$. The gray lines shown in Figure~\ref{Fig:ngalr} are the median values of $f_{\mu}(R)$ for galaxies
in different radial bins for each mock cluster, and the red line is the median $f_{\mu}(R)$ for the cluster sample.
It should be mentioned that the differences of the median $f_{\mu}$ are less than $2\%$ at radius $R>0.15h^{-1}\hbox{Mpc}$ when changing
the amplitude of $M$--$c$ relation (Equation \ref{eq:cmp}) from $A_0 = 2$ to 6. Thus the estimated median depletion fraction is not very sensitive to
the input amplitude of the $M$--$c$ relation for the fitting range $R >0.15h^{-1}\hbox{Mpc}$.
With $f_{\mu}$, the interloper fraction $f_{\rm I}$ for each source galaxy can then be estimated
by Equation (\ref{eq:mbias}) using the value of $f_{\rm n}$ shown by the black solid line in Figure~\ref{Fig:ngalr}.

To mimic the dilution effect in simulations, for each source galaxy, we randomly assign it a value of $p$ in the range of $0$ to $1+f_{\rm I}(R)$. Only galaxies with $p\le1$ are taken as source galaxies and given the lensing signals according to the step (4) described above.
For galaxies with $p>1$, they are regarded as cluster members, and therefore no lensing signals are given. In this way, we generate
a mock catalog with the interloper rate estimated by taking into account the mass-dependent magnification bias.
For comparison, we also generate another mock catalog by assuming the interloper fraction $f_{\rm I}=f_{\rm n}(z_{s,d}>0.1)$,
that is, assuming that the distribution of background galaxies is uniform without including the lensing magnification bias.

We then have six sets of mock simulations in total, that is, four sets of simulations for all $220$ clusters without dilution effect
and two sets of simulations only for $158$ isolated clusters with dilution effect.
We perform NFW fitting for each of the mock clusters following the same procedures as for the real observational analyses
and study the $M$--$c$ relation subsequently for each realization. In this way, using Monte Carlo simulations, we can
directly investigate the bias effects caused by shape noise, center offset, redshift uncertainty, and our NFW fitting method.

\section{Results}

In this section, we present the results of $M$--$c$ relation analyses. For both the observational sample and each realization of simulations,
we only consider clusters with the lensing-derived mass in the range of $M_{14}=10^{14}\hms$ to $8M_{14}\simeq10^{14.9}\hms$,
and we divide them into four equal bins in $\log M$ space. In each mass bin, we can obtain the median value of the concentration parameter from the
clusters therein.

Figure \ref{Fig:cfhtmcr} shows the $M$--$c$ relation obtained from our observational analyses (top panels) and from
our Monte Carlo simulations (bottom panels). From left to right, the first column presents the results based on the direct NFW fitting
for individual clusters without priors on mass and without accounting for the effect of center offset.
The second column is for the results with the mass prior $P(M|\lambda)$ according to the mass--richness relation for redMaPPer clusters,
as explained in Section 3.2.1. The last column plots the results for $158$ isolated clusters, including both the mass prior and
the center-offset distribution $P(R_{\rm off})$. Note that when accounting for the center-offset effect,
we introduce an additional fitting parameter $R_{\rm off}$. For each paired cluster, there are then a total of six fitting parameters.
With CFHTLenS lensing data, we cannot derive reliable constraints for them.
Therefore we only show the results for isolated clusters in the rightmost panels.
In the following, we explain the results shown in each panel in detail.

\subsection{$M$--$c$ Relations from Monte Carlo Simulations}

\begin{figure*}
  \centering
  \includegraphics[width=0.9\textwidth]{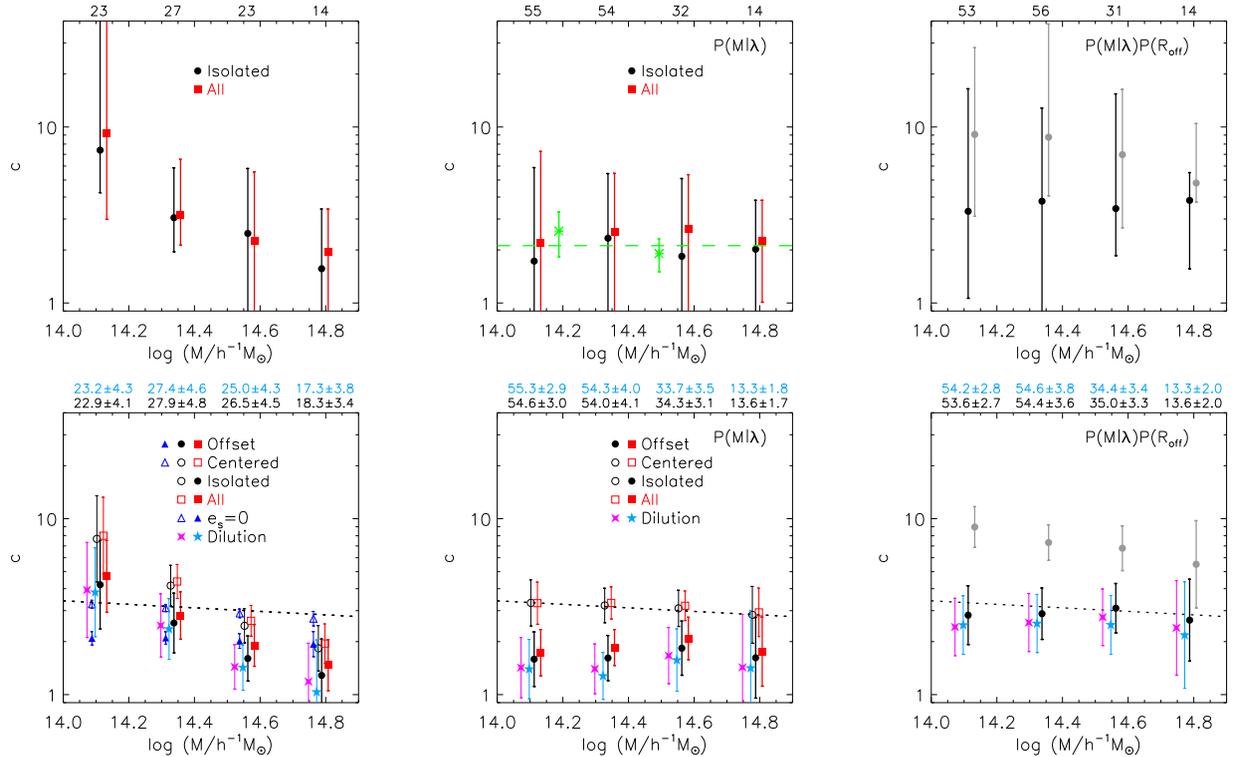}\\
  \caption{$M$--$c$ relations from CFHTLenS (top panels) and the corresponding Monte Carlo simulations (bottom panels). From left to right, the first column presents the results based on the direct NFW fitting without priors on mass and center offset. The second column is for results with the mass prior $P(M|\lambda)$ based on the mass--richness relation for redMaPPer clusters. The last column plots the results for isolated clusters, including both the mass prior and the center offset distribution $P(R_{\rm off})$, where by default the redMaPPer-derived centering probability $P_{\rm cen}$ is used. The number of isolated clusters in each mass bin is shown at the top of each panel. Black and sky-blue numbers respectively correspond to filled black circles and sky-blue five-pointed stars.
  Top panels: the filled black circles and red squares show the median concentrations, respectively, for isolated clusters and all the clusters including paired ones. The gray circles in the top right panel show the results with an unreasonable assumption of $P_{\rm cen}=0.3$ for all of the isolated clusters. The error bars around circles and squares show the range of the $25\%-75\%$ percentile. The green dashed line in top middle panel indicates the concentration $c=2.11$ as shown in Figure \ref{Fig:ngalr}. Green asterisks with error bars show the estimated concentrations based on number density distribution $\hat{n}_m$ for low- ($<34$) and high-richness ($>34$) clusters.
  Bottom panels: symbols indicate the medians of median concentrations for 100 realizations of the corresponding simulations. The error bars show the corresponding scatters of medians. Specifically, the filled and open symbols are respectively for the simulations with and without center offset. Circles and squares are respectively for isolated clusters and all clusters including pairs.
  Triangles denote the cases for pure lensing signals. The gray circles in the bottom right panel present the results with $P_{\rm cen}=0.3$.
Star polygons present the results for simulations with interlopers from member galaxies. Magenta four-pointed and sky-blue five-pointed stars
are respectively for simulations with the dilution effect estimated from $f_{\rm I}=f_{\rm n}$ and $f_{\rm I}=(1+f_{\rm n})/(1+f_\mu)-1$.
The dotted line shows the fiducial input of the $M$--$c$ relation in our Monte Carlo simulations. Note that in this paper
the overdensity parameter for all mass definitions is $\Delta=200$ relative to critical density $\rho_{\rm crit}$.
  }\label{Fig:cfhtmcr}
\end{figure*}

To see clearly the influences from different effects, we first focus on the $M$--$c$ relations from Monte Carlo simulations
shown in the bottom panels of Figure \ref{Fig:cfhtmcr}. We show the fiducial input of the $M$--$c$ relation with the dotted line for comparison. The filled and open symbols are, respectively,
for simulations with and without center offset. That is, for results shown by open symbols, we center our simulated clusters
at the positions of the corresponding central galaxies. For filled symbols, the cluster centers are randomly assigned around the
central galaxies according to the distribution of Equation (\ref{eq:proff}), as described in the Section 4.
Circles and squares are, respectively, for isolated clusters only and for all clusters including paired ones.
Triangles denote the cases without intrinsic shape noises. The other two sets of simulation results
for isolated clusters including the dilution effect are denoted by star polygons.
Magenta four-pointed stars and sky-blue five-pointed stars are respectively for the median concentrations
from the set of simulations with the interloper fraction $f_{\rm I}=f_{\rm n}$ and $f_{\rm I}=(1+f_{\rm n})/(1+f_\mu)-1$. The mass dependence
of $f_{\mu}$ is taken into consideration (see Section 4).
We note that, for each set of simulations, we have 100 realizations. For each realization, we can obtain the median $c$ in each mass bin.
The symbols shown in the bottom panels indicate the medians of median concentrations for the 100 realizations of the corresponding simulations,
and the error bars are for the dispersion of the median concentrations within the 100 realizations.
The number of isolated clusters in each mass bin is written out at the top of each panel,
with the black ones for the fiducial simulation with offset and noise and the sky-blue numbers for the simulation set
with offset, noise, and the interlopers calculated by accounting for the magnification bias.
Note that these numbers are the average values over the
100 realizations for each simulation set and therefore are not integers.

The bottom left panel presents the derived $M$--$c$ relations from direct profile fitting for individual clusters without mass priors $P(M|\lambda)$.
Also, the fitting for a cluster is done using the position of its central galaxy as the center of the cluster.
We can see immediately that, for each pair of symbols, the filled ones are systematically lower than the open ones.
We remind readers here that for the filled ones the true centers of the simulated clusters are not at the positions of the central galaxies,
but for the open ones, they are by setting. Therefore the results here show that the wrong center usage tends to bias the concentration estimation
to lower values, and the offset effect is rather significant.

The dilution effect can further lower the derived concentrations, but it is minor
compared to the off-center effect for our cluster sample.
By comparing the five-pointed stars with the four-pointed stars, we see that the amplitude of the negative bias is increased about
$4\%$ on average, with larger bias for massive clusters,
when the lensing magnification effect on the spatial distribution of background galaxies is considered.

It is also seen obviously that the slope of the $M$--$c$ relation is steeper for the cases with shape noise than for those without the noise
with $\boldsymbol {e_s}=0$. This phenomenon has been studied in detail in \citet{2014ApJ...785...57D}. We show that,
because of the $(M, c)$ degeneracy in terms of the reduced shear signals, the errors in $(M, c)$ are strongly correlated with a higher mass
compensated for by a lower $c$. The larger the noise level is, the larger the errors in $(M, c)$ are. The error in mass leads to
misassignments of clusters to the wrong mass bins. Consequently,  the larger mass bins tend to have lower $c$ and vice versa. The fact that
the degeneracy direction of $(M, c)$ is steeper than the true $M$--$c$ relation results in an apparently steeper $M$--$c$ relation than the underlying
true one. For the case without center offset and shape noise shown by the open triangles, the derived $M$--$c$ relation is
nearly the same as the input one indicated by the dotted line, which demonstrates that our analysis method itself does not introduce
significant biases to the derived $M$--$c$ relation.

The bottom middle panel displays the simulation results with the mass prior $P(M|\lambda)$ according to the mass--richness relation.
It is seen that in comparison with the results shown in the bottom left panel,
the slope bias in the derived $M$--$c$ relation is significantly reduced because of the reduction of scatter on the
mass estimate, which in turn largely breaks the degeneracy between mass and concentration.
The open symbols now are very close to the input dotted line. For the solid ones, the slope is about the same as the input one, but the
amplitude is low because of the offset effect and the dilution effect.
We can also notice that the amplitude of concentration including paired clusters is somewhat higher than isolated clusters in the case
with off-center effects (red solid squares versus the black solid circles). It is due to the different fitting ranges for profile studies
for isolated and paired clusters, with the latter being larger. Therefore, the weak lensing analyses of paired clusters are less affected by the off-center effect than that of isolated clusters.
On the other hand, for the case without center offset, the results for isolated clusters and for all clusters including the paired ones
are about the same (open black and red symbols).

The bottom right panel presents the simulation results with the mass prior and accounting for the off-center effect
by introducing $P(M|\lambda)P(R_{\rm off})$ in the cluster profile fitting, as shown in Equation (\ref{eq:probmc}).
The move upward of the black solid circles manifests that the negative bias caused by center offset can be largely reduced
by properly considering the offset distribution.
The residual biases are small, with the values of $\sim[-17,-9,4,-4]\%$ from the low- to high-mass bins,
which can be attributed to the fact that our method, including the mass prior and the center-offset probability in the analyses,
may not be able to deal perfectly with the cases with very large noise and center offset, especially for low-mass clusters with low lensing signals.
We emphasize again that the applicability of our analytical method here relies on the information of $P_{\rm cen}$
given to each central galaxy candidate, as redMaPPer does.
To demonstrate the importance of a correct estimate of $P_{\rm cen}$ in weak lensing cluster studies,
we show in gray circles the results with $P_{\rm cen}=0.3$. It is seen that
with this unreasonably low centering probability, the effect of center offset on profile studies is overcorrected,
and subsequently the derived $M$--$c$ relation is biased to high values in amplitude.

The magenta and the sky-blue symbols in the bottom right panel show the results including the
dilution effect without and with the lensing magnification bias, respectively.
The differences between magenta stars and black solid circles are nearly constant ($\sim [11, 10, 11, 9]\%$) in
different mass bins because the same median interloper fraction $f_{\rm I}=f_{\rm n}$ is used for each cluster in simulations.
By comparing the sky-blue symbols with the black filled circles of the fiducial case, we can
see that the dilution effect with the mass-dependent lensing magnification bias introduces a mass-dependent negative bias with the values of $\sim [10, 11, 20, 17]\%$ from
low- to high-mass bins.
Furthermore, with the residual bias of the black filled circles discussed above,
we find from our mock simulation studies that the total bias for the sky-blue symbols compared to the input $M$--$c$ relation (dotted line)
is $\approx [-27, -20, -16, -21]\%$ for the four mass bins. In Section 5.3, we will use these biases to calibrate the observational result.

In summary of this subsection, the Monte Carlo simulation studies shown here demonstrate well the feasibility of our approaches to derive the $M$--$c$ relation
by analyzing clusters individually with careful priors on mass and on centering probabilities. In the following, we will show the results from CFHTLenS observations.

\begin{figure*}
  \centering
  \includegraphics[width=0.9\textwidth]{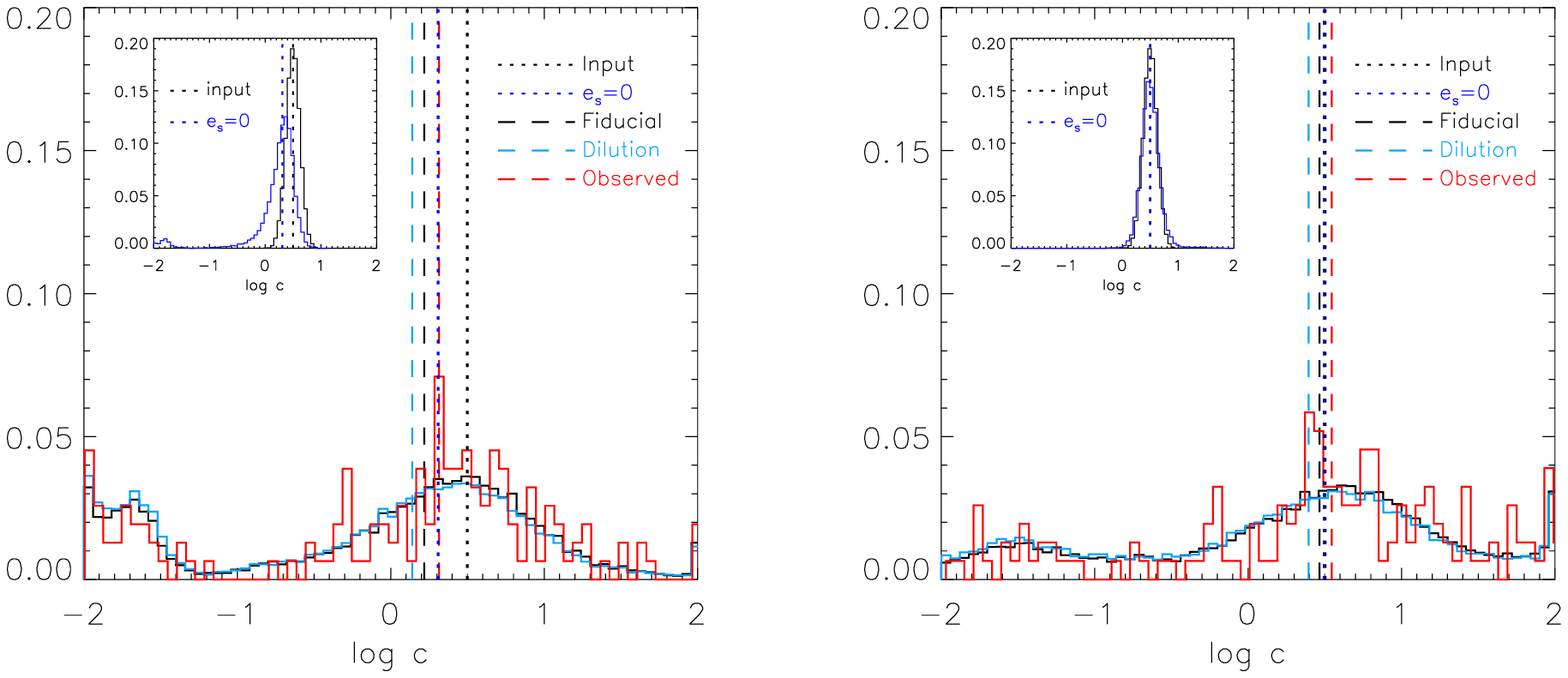}\\
  \caption{Concentration distribution in log space. Left and right panels show the results with prior $P(M|\lambda)$ and
$P(M|\lambda)P(R_{\rm off})$, respectively. The black and blue distributions in subpanels respectively correspond
to the original input concentrations and the recovered concentrations with certain priors. The median values are shown with
vertical dotted lines. The black and sky-blue histograms in the main panels respectively display the concentration distributions
for fiducial simulations and that with the dilution effect accounting for the lensing magnification bias.
Red is for the observational result. Vertical dashed lines show the corresponding medians. Note: the overdensity parameter $\Delta=200$ is adopted.
}\label{Fig:siglgc}
\end{figure*}

\subsection{$M$--$c$ Relation from CFHTLenS}

The top panels in Figure \ref{Fig:cfhtmcr} exhibit the observational results of the $M$--$c$ relation using CFHTLenS. Red squares are for all of the clusters,
including the paired ones. Filled black circles are for isolated clusters. The error bars indicate the percentile range of $25\%-75\%$ around the
median values. This quotation of percentiles for dispersion is mainly due to the wide and complicated distribution of observed concentrations.
We can see clearly that the trend from the left to right panels is very similar to the simulation results shown in the bottom panels.
This demonstrates that the shape noise can indeed make the apparent $M$--$c$ relation steeper than
the underlying true $M$--$c$ relation (top left). With proper mass priors in profile fittings, this steepening can largely be removed (top middle).
To further include the centering probability in the analyses, the underestimation in $c$ due to center offsets is
improved significantly (black solid symbols in the top right panel). We also show using an example of $P_{\rm cen}=0.3$ that an unreasonably low centering probability can result in
much higher concentrations (gray symbols in the right panel). As explained in Section 5.1, we only show the results for isolated clusters in the top right panel
because the data cannot give reliable constraints for paired clusters containing six fitting parameters.
There are in total $154$ isolated clusters in the mass range from $M_{14}=10^{14}\hms$ to $8M_{14}\simeq10^{14.9}\hms$ (right panel).
Comparing to the mock simulation results shown in the bottom panels, we see that the number of isolated clusters in each mass bin in each panel
is very consistent with that of the corresponding simulation results, even for that without priors on mass and center offset (left panels).
This consistency demonstrates the feasibility of the usage of the mass richness relation for mass priors, in which it is used to
assign a mass to a cluster in our mock simulations.

In the top middle panel, we also show the results by fitting the member galaxy distribution to an NFW profile (green).
Green asterisks present the concentrations derived from the stacked number density of galaxies $\hat{n}_m$ for low- ($<34$) and
high-richness ($>34$) clusters, respectively. The green dashed line indicates the $c=2.11$ obtained from fitting the NFW profile
to the full stacked number-density profile, as shown in Figure \ref{Fig:ngalr}.
We see that the derived concentrations based on the number distribution of member galaxies are in accord with the results from weak lensing analyses,
and the center-offset effect is significant.

Besides the $M$--$c$ relation, we also analyze the distribution of weak-lensing-derived concentrations for our cluster samples
with weak-lensing-derived mass in the range of $M_{14}$ to $8M_{14}$.
The results are shown in Figure \ref{Fig:siglgc}. Because of the strong dependence of the derived concentration on mass for direct profile fitting,
we only present the $\log c$ distribution with the priors of $P(M|\lambda)$ (left) and $P(M|\lambda)P(R_{\rm off})$ (right), respectively.
The black and blue distributions in the subpanels correspond respectively to the original input concentrations and the recovered concentrations from our
Monte Carlo realizations with $\boldsymbol e_s=0$. The median values are indicated by vertical dotted lines.
The black histograms in the main panels display the results from our fiducial Monte Carlo simulations. The sky-blue ones show the results
from the simulations, including the dilution effect with the interloper fraction $f_{\rm I}$ estimated by accounting for
the excess fraction $f_{\rm n}$ and the lensing depletion fraction $f_\mu$.
The observational results are shown in red. Vertical dashed lines show the corresponding medians.

We first investigate the results shown in the subpanels of Figure \ref{Fig:siglgc} without shape noise.
In the left one, we see that the results from the noiseless simulations are systematically
biased to lower values than the input ones because of the effect of center offset. By properly accounting for the offset effect,
this bias can be eliminated, as seen from the right subpanel.
This shows again that our fitting method is reliable without introducing artificial biases.

For the results presented in the main panels of Figure \ref{Fig:siglgc}, it is seen clearly that the general behaviors
of the distributions from observations (red) and from our mock Monte Carlo simulations are very similar. The high shape noise level
of the CFHTLenS and the center-offset effect lead to a wide distribution for
the derived concentration parameter.
In the left panel, there is a significant bump at the low-concentration end that is caused
by the cases with too-large center offsets and high noise. The bump decreases significantly in the right panel when
the offset effect is accounted for in the analyses. On the other hand, the influences of the large noise level and center offset can be
correlated. Therefore, unlike the results from simulations without noise (subpanels), our method
taking into account the center-offset probability cannot fully recover the intrinsic $M$--$c$ relation with such large noise.
However, the residual negative bias is small, as indicated by the vertical black dashed line in the right panel.
By comparing the sky-blue dashed lines with the black dashed lines, it is seen clearly that the dilution effect introduces an additional negative bias.
The results here are in full accord with the results shown in Figure~\ref{Fig:cfhtmcr}.

By comparing the red and sky-blue histograms, we see that the observational distribution is very similar
to that of mock simulations. On the other hand, the median value of $\log c$ from observations as indicated by the red dashed line
is somewhat larger than the result of mock simulations shown by the sky-blue dashed line.
In the next subsection, we will compare our observational result with other studies.

\subsection{Comparisons with Other $M$--$c$ Relation Studies}

\begin{figure}[!ht]
  \centering
  \includegraphics[width=0.45\textwidth]{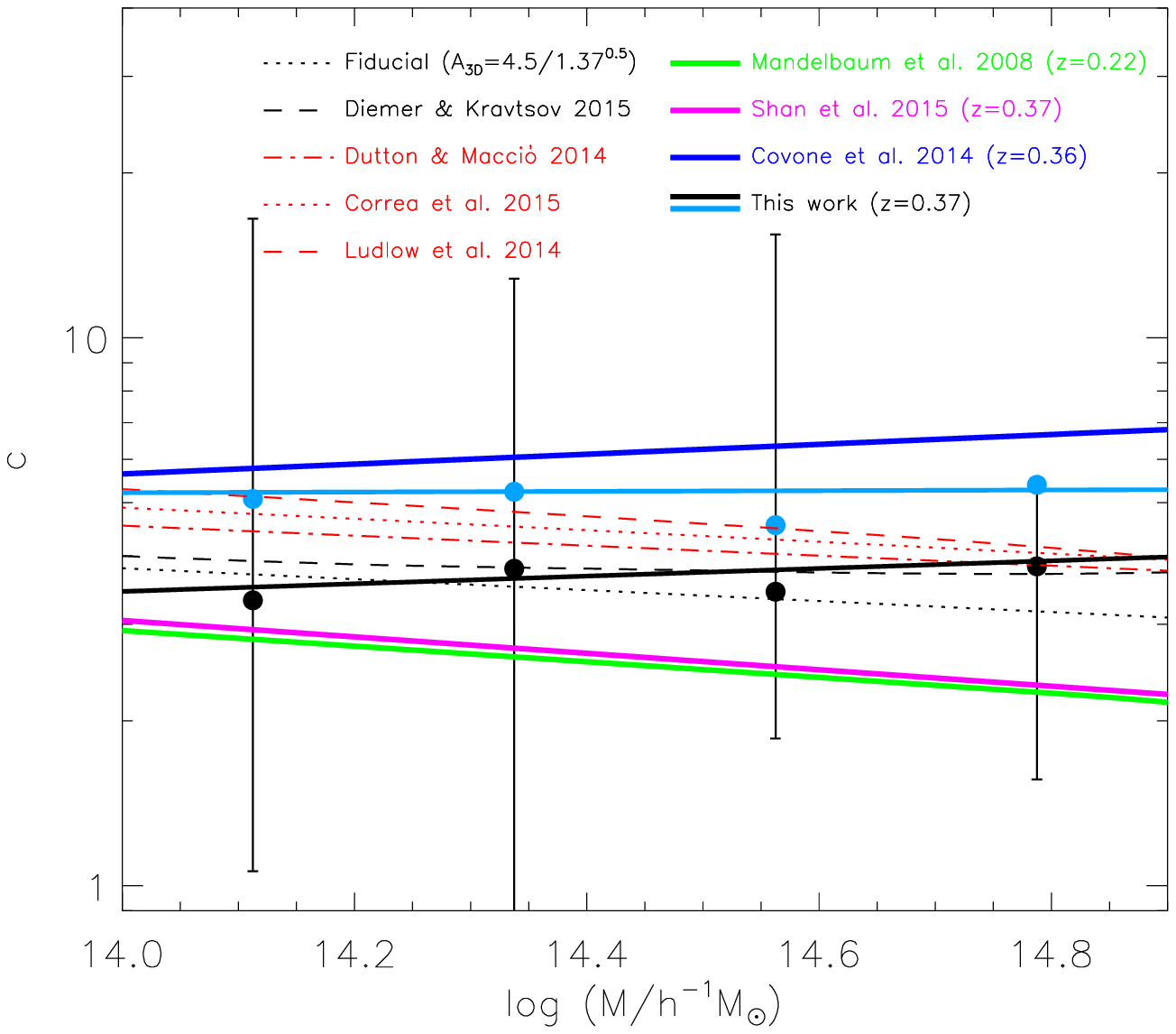}\\
  \caption{Comparisons with previous $M$--$c$ relations. Thin lines correspond to $M$--$c$ relations from numerical simulations
in {\it Planck} cosmology at redshift $z=0.37$. Black and red thin lines are respectively for full sample and relaxed clusters.
Among them, the black dotted line indicates the 3D fiducial relation inferred from the \citet{2008MNRAS.390L..64D} $M$--$c$ relation.
Thick solid lines are for the observed relations. Green, magenta, and blue thick lines present the 2D observational results from
\citet{2008JCAP...08..006M}, \citet{2015arXiv150200313S}, and \citet{,2014ApJ...784L..25C}, respectively. Filled black circles
correspond to the median concentrations (the same as the filled black circles in the top right panel of Figure \ref{Fig:cfhtmcr})
from NFW fitting without bias correction of dilution and projection effects. Error bars present the $25\%$ and $75\%$ percentiles
around the median values. The black thick line shows the fitting to filled black circles with $A(z=0.37)=3.45^{+1.58}_{-0.89}$
and $\alpha=0.07^{+0.28}_{-0.32}$. Sky-blue circles show the 3D amplitude of concentrations predicted from filled black circles
enhanced by $1/(1-10\%)/(1-b)$, where the factor $10\%$ and $b\simeq[27,20,16,21]\%$ are respectively
for the $\sim 10\%$ projection effect and the residual bias from our analyses plus the dilution bias.
The sky-blue thick line presents the fitted relation to sky-blue circles with $A(z=0.37)=5.21^{+2.38}_{-1.35}$
and $\alpha=0.006^{+0.27}_{-0.32}$. Note: the overdensity parameter relative to critical density is $\Delta=200$ for all of the $M$--$c$ relations here.
  }\label{Fig:mccomp}
\end{figure}

In this subsection, we compare our observational result from CFHTLenS with other observational studies on $M$--$c$ relations
and with results from numerical simulations. Different definitions of concentration and mass can affect the trend of the $M$--$c$ relation
\citep{2013arXiv1303.6158M}. For consistency, we only choose the simulation results for which the concentration is defined from
direct density profile fitting \citep{2014MNRAS.441.3359D,2014MNRAS.441..378L,2015MNRAS.452.1217C,2015ApJ...799..108D}.
For other observational studies, we choose three weak-lensing-derived $M$--$c$ relations based on the stacking method
to be compared with \citep{2008JCAP...08..006M,2014ApJ...784L..25C,2015arXiv150200313S}.
We compare with these stacked studies because, in our analyses, we apply a mass prior based
on redMaPPer cluster richness to suppress the effect of the correlated errors of $M$ and $c$ on the weak-lensing-derived $M$--$c$ relation.
This is similar in essence to the stacked analyses where the stacking is based on the richness of clusters or the luminosity/stellar mass
of central galaxies. The results are shown in Figure \ref{Fig:mccomp}.

The thin lines are the $M$--$c$ relations at $z\approx 0.37$ from a few numerical simulations for {\it Planck} cosmology.
Specifically, the red thin lines show the $M$--$c$ relations for ``relaxed'' halos given
by \citet{2014MNRAS.441.3359D} (dash-dotted), \citet{2015MNRAS.452.1217C} (dotted), and \citet{2014MNRAS.441..378L} (dashed).
The black thin lines are for the $M$--$c$ relations for ``all simulated'' dark matter halos.
The dashed black thin line presents the $M$--$c$ relation from \citet{2015ApJ...799..108D}. The 3D relation with $A_{\rm 3D}=4.5$ at $z=0.37$ from \citet{2008MNRAS.390L..64D} adjusted to {\it Planck} cosmology
is shown by the black dotted line, which is the one used in our mock simulations (with an additional correction of $\sim -10\%$
for the projection effect).
We can see that the concentration parameters of relaxed halos are systematically higher than that
including unrelaxed halos. In the narrow mass range considered here, the $M$--$c$ relation from numerical simulations can be well approximated by
a power-law relation with the power index $\alpha\sim -0.1$. The exception is the result from \citet{2015ApJ...799..108D}, which shows a slightly positive slope
at the very high mass end.

The thick solid lines in Figure \ref{Fig:mccomp} are for the observational results. The green and blue thick lines represent the results
from \citet{2008JCAP...08..006M} and \citet{2014ApJ...784L..25C}, respectively, for which the reference redshift and cosmology in their work are used.
The magenta thick line shows the updated result from \citet{2015arXiv150200313S} with $A_0=6.77$, $B=0.67$, $\alpha=-0.15$,
and $M_{\rm p}=2\times10^{12}h^{-1}\msun$ for {\it Planck} cosmology. Note that for the result of \citet{2008JCAP...08..006M} (green),
the original mass definition is with respect to the mean density of the universe. Here we have converted it to be consistent with our mass definition
with respect to the critical density $\rho_{\rm crit}$ of the universe \citep{2003ApJ...584..702H,2012MNRAS.424.1244F}.
Filled black circles are the median concentrations (the same as the filled circles in the top right panel of Figure \ref{Fig:cfhtmcr}) from our
weak lensing analyses without accounting for the dilution effect. Error bars present the $25\%$ and $75\%$ percentiles around the median values.
The black thick line shows the best-fit relation to the black filled circles based on Equation \ref{eq:cmp}.
We find $A(z=0.37)=3.45^{+1.58}_{-0.89}$ and $\alpha=0.07^{+0.28}_{-0.32}$ where the errors are estimated from bootstrap resampling.
We note that with the applied mass priors in our weak lensing analyses,
the steepening of the $M$--$c$ relation due to the correlated errors of $M$ and $c$ is eliminated, as shown in Figure~\ref{Fig:cfhtmcr}
(middle panels versus left panels). Thus here our $M$--$c$ relation fitting is done directly without considering the error correlations.

It is known that weak lensing analyses are directly related to the 2D projected density profile of a cluster.
As we discussed in Section 4, compared to the 3D concentration parameter,
there is a $\sim 10\%$ negative bias for the 2D weak-lensing-derived $c$.
Furthermore, our Monte Carlo simulations reveal the existence of bias from the dilution effect and the imperfect treatments of our method
for cases with very large center offsets and high noise.

Therefore, for a better comparison with 3D results from numerical simulations,
we correct for these negative biases by multiplying the 2D medians, shown as the black filled circles in Figure \ref{Fig:mccomp},
by a factor of $1/(1-10\%)/(1-b)$. Here, the number $10\%$ accounts
for the bias from the pure projection effect. The factor $b$ describes the additional bias
with $b\simeq[27,20,16,21]\%$ from the low- to high-mass bins calibrated from our mock studies
by calculating the ratio of the sky-blue values to the corresponding values on the dotted line in the bottom right panel of Figure \ref{Fig:cfhtmcr}.
The sky-blue circles in Figure \ref{Fig:mccomp} present the corrected 3D $M$--$c$ relation from our analyses. The sky-blue line is for the
fitted $M$--$c$ relation with $A(z=0.37)=5.21^{+2.38}_{-1.35}$ and $\alpha=0.006^{+0.27}_{-0.32}$.
We see that our derived $M$--$c$ relation is in agreement with most of the numerical simulation results with the power-law slope
close to zero. The normalization is somewhat higher than that given by simulations.

In comparison with the other three observational results, the normalization of our derived $M$--$c$ relation
lies in the middle range of them. We showed in the right panels of Figure \ref{Fig:cfhtmcr} that an unreasonably low $P_{\rm cen}$
assumed in the analyses can overcorrect the center-offset effect and lead to a significantly positive bias to the normalization of the $M$--$c$ relation.
In \citet{2014ApJ...784L..25C}, they also used CFHTLenS but for stacked weak lensing analyses. The cluster catalog is taken from \citet{2012ApJS..199...34W},
in which there is no specific centering probability information available for individual clusters. Therefore, in their analyses,
they assumed a large constant fraction ($>50\%$) of center-offset halos. This may explain the high $M$--$c$ relation they obtained.
Note that for redMaPPer clusters the average $P_{\rm cen}$ is $\sim 0.92$, and the fraction of clusters with misidentified optical centers is only $\sim 8\%$.

The results of \citet{2008JCAP...08..006M} and \citet{2015arXiv150200313S} are very consistent with each other both in slope and in normalization.
Our $M$--$c$ relation is consistent with theirs in slope, but the normalization is considerably higher.
In \citet{2008JCAP...08..006M}, the weak lensing data is from the Sloan Digital Sky Survey \citep{2000AJ....120.1579Y,2005MNRAS.361.1287M},
and the foreground lenses cover from galaxies to groups to maxBCG clusters \citep{2001AJ....122.2267E,2006MNRAS.368..715M,2007ApJ...660..239K}.
In their stacked weak lensing analyses,
they argued that the center-offset effect can be controlled by limiting the fitting range to relatively large radii of a cluster
in weak lensing analyses.
As a test, we use our mock simulations containing center offsets but without noise to analyze how the lower limit of the radial fitting range
can suppress the off-center effect. We find that the bias from the center-offset effect is negligible only for a very large lower limit of $1\hmpc$
with respect to the chosen central galaxies. With the lower cuts of $0.5\hmpc$ and $0.15\hmpc$, the systematic bias can
still reach to a level of $\sim-12\%$ and $\sim-31\%$, respectively.
We also note that the choice of a suitable lower limit for the radial fitting range to
control the offset effect depends on the offset distribution of the cluster sample used in the analyses.
Therefore the negative bias resulting from the center offset can be hardly eliminated fully by this treatment.

\citet{2015arXiv150200313S} used the CFHT Stripe 82 Survey for weak lensing analyses, and
the foreground clusters and groups were taken from the redMaPPer cluster catalog and from the LOWZ/CMASS galaxies of the Sloan Digital Sky Survey-III
Baryon Oscillation Spectroscopic Survey Tenth Data Release (SDSS-III BOSS DR10) \citep{2011ApJS..193...29A,2011AJ....142...72E,2013AJ....145...10D}.
In their analyses, the off-center effect is accounted for. On the other hand, the results might suffer from significant dilution effects from
member galaxies due to the inaccurate photo-$z$ estimate for the source galaxies
used in their studies (\citealt{2015arXiv150200313S}).

In summary, our obtained $M$--$c$ relation for redMaPPer clusters using CFHTLenS weak lensing data is consistent with numerical simulation results within
error bars. The normalization is somewhat higher than that from simulations. Given the relatively large scatters from different observational
studies, our resulting $M$--$c$ relation falls in the middle range of the three weak lensing analyses compared here. Different lensing data and foreground cluster catalogs and different analyzing methods can all contribute to the differences.

\section{Summary}

Weak lensing analyses play important roles in probing the mass distribution of clusters of galaxies from inner to outer parts.
CFHTLenS covers about $150\deg^2$, within which there are more than $200$ redMaPPer clusters.
We therefore can study the $M$--$c$ relation for this large sample of clusters using weak lensing data all from the same survey
with similar observational conditions.

Comparing to targeted weak lensing cluster observations \citep[e.g.,][]{2010PASJ...62..811O,
2012MNRAS.420.3213O,2015arXiv150704493O,2015arXiv150704385U}, CFHTLenS is relatively shallow, with a
typical source galaxy number density of $n_{\rm g}\sim 8\hbox{ arcmin}^{-2}$.
Thus the shape noise is large for individual cluster analyses, which in turn results in a significant
steepening effect on the weak-lensing-derived $M$--$c$ relation, due to the correlated errors of $M$ and $c$.
Furthermore, at this large noise level, the error distributions of $M$ and $c$ cannot be well modeled as Gaussian ellipses.
Thus the Bayesian methods developed to extract the underlying $M$--$c$ relation by including correlated Gaussian errors of $M$ and $c$
\citep{2014ApJ...785...57D, 2015arXiv150704493O, 2015arXiv150704385U} are not suitable for CFHTLenS observations.
Stacked studies can suppress the noise effect \citep[e.g.,][]{2008JCAP...08..006M, 2014ApJ...784L..25C}.
However, complications to deal with center offsets and so on may enter into the stacked analyses.

In this paper, we perform weak lensing analyses using CFHTLenS for individual redMaPPer clusters.
To overcome the large noise effect, we employ an external mass prior based on
the mass--richness relation of the cluster sample. Including further the
the centering probability $P_{\rm cen}$ given to each central galaxy candidate
by the redMaPPer algorithm, we introduce a statistical method for individual profile fitting
that can reduce significantly the effects from noise and the center offset.
Using Monte Carlo simulations, we demonstrate the feasibility of the method.

To show the influences of different effects and how our proposed method works, we analyze each individual cluster in different ways: direct fitting
without any priors, applying a mass prior only, and applying a mass prior and accounting for the effect of center offset.
For direct-fitting analyses, the derived $M$--$c$ relation is significantly steeper than that expected from simulations, due to the large shape noise,
consistent with our previous studies \citep{2014ApJ...785...57D}.
By adding a mass prior based on the mass--richness relation for redMaPPer clusters
obtained from the abundance matching, we show that the degeneracy between mass and the concentration parameter can be broken.
Consequently, the steepening of the $M$--$c$ relation induced by the shape noise is nearly eliminated,
and its power-law index is very close to that from simulations. The amplitude,
however, as revealed from our mock simulation studies, is clearly low, which is mostly due to the center-offset effect.
We demonstrate that this can be corrected for by adding the information of centering probability for each individual cluster into the analyses.
The residual bias is small at the level of $\sim -6\%$.

The dilution effect from cluster members and foreground galaxies before clusters can lead to an additional negative bias to
the weak-lensing-derived concentration parameter. We note that the photo-$z$ distribution for each source galaxy given by CFHTLenS
is very helpful in reducing the dilution effect (Equation [\ref{eq:criden1}]).
However, the source galaxies selected only based on their photo-$z$ values can still be contaminated because of the uncertainties of the photo-$z$ estimation.
We find a residual excess fraction of $f_{\rm n} \sim 7\%$ from member galaxies by investigating the enhanced number-density distribution of
the selected CFHTLenS source galaxies in cluster regions. By taking into account the lensing magnification bias to
the number density of background galaxies, the interloper fraction $f_{\rm I}$ in the inner part of clusters has a median value of $\sim12\%$ for our cluster sample.
The corresponding dilution effect leads to a $\sim 15\%$ negative bias to the amplitude of the $M$--$c$ relation, according
to our Monte Carlo simulation studies. The effect depends on the mass of clusters because of the mass-dependent lensing magnification bias.

With corrections for the bias from the dilution effect and the residual bias calibrated from our mock simulation analyses, and
also the $\sim 10\%$ negative bias from the projection effect,
we obtain the 3D $M$--$c$ relation with $\alpha= 0.006^{+0.27}_{-0.32}$ and $A(z=0.37)=5.21^{+2.38}_{-1.35}$ for $M_p=M_{14}$ from our studies.
The result is in agreement with that from N-body simulations, with a weak tendency that the observed amplitude $A$
is somewhat higher.

Our studies here show that even with surveys that have a relatively large shape noise,
an unbiased $M$--$c$ relation can still be derived from a large sample of clusters with their profiles constrained
individually from weak lensing analyses. Future weak lensing surveys will cover much larger areas, and the number of
clusters therein will increase by orders of magnitude. We then expect much better constraints on the $M$--$c$ relation
from weak lensing cluster analyses. Furthermore, the shape noise from future deeper surveys will be considerably lower, and the cluster centers
can be constrained better. Then the Bayesian methods, taking into account the correlated Gaussian errors
of $M$ and $c$, will be applicable to extract the unbiased $M$--$c$ relation without relying on external mass priors.

\acknowledgments

We thank the referee very much for the encouraging and detailed comments that help improve the paper considerably.
W.D. and Z.H.F. acknowledge the support from NSFC of China under the grants 11333001, 11173001, and 11033005.
H.Y.S. acknowledges the support by a Marie Curie International Incoming Fellowship within
the 7th European Community Framework Programme, and NSFC of China under grant 11103011.
L.P.F. acknowledges the support from NSFC grants 11103012 and 11333001 and from Shanghai Research grant 13JC1404400 of STCSM.
J.P.K. acknowledges support from the ERC advanced grant LIDA and from CNRS. G.B.Z. is supported by the 1000 Young Talents program in China. Z.H.F. and G.B.Z. are also supported by the Strategic Priority Research Program ``The Emergence of Cosmological Structures'' of the Chinese Academy of Sciences Grant No. XDB09000000.

\bibliography{cfhtmcrev}
\end{document}